\documentstyle[12pt,amsmath,amssymb,epsf,epic]{article}

\newcommand{\h}{{\cal H}}
\newcommand{\s}{{\cal S}}
\newcommand{\e}{{\rm e}}
\renewcommand{\c}{{\cal C}}
\newcommand{\bbbone}{\mathchoice {\rm 1\mskip-4mu l} {\rm 1\mskip-4mu l}
{\rm 1\mskip-4.5mu l} {\rm 1\mskip-5mu l}}
\newcommand{\spec}{{\rm spec}(M)}
\newcommand{\ee}{{\rm e}}
\newcommand{\ch}{{\cal H}}
\newcommand{\cm}{{\mathfrak M}}
\newcommand{\scalprod}[2]{\left\langle {#1}, {#2}\right\rangle}
\newcommand{\p}{{\cal P}}
\renewcommand{\i}{{\rm i}}
\newcommand{\av}[1]{\left\langle {#1} \right\rangle}
\newcommand{\fer}[1]{(\ref{#1})}
\newcommand{\pbar}{{\, {\!\overline P}} }

\newcommand{\E}{{\mathbb E}}
\renewcommand{\d}{{\rm d}}
\newcommand{\U}{{\mathcal U}}

\newcounter{resultcounter}

\newtheorem{lem}[resultcounter]{Lemma}
\newtheorem{prop}[resultcounter]{Proposition}
\newtheorem{trace}[resultcounter]{Corollary}
\newtheorem{thm}[resultcounter]{Theorem}

\begin{document}

\title{Quantum Measurements of Scattered Particles} 
\author{ Marco Merkli,\footnote{Department of Mathematics and Statistics, Memorial University of Newfoundland, St. John's, NL
Canada, A1C 5S7; merkli@mun.ca; Supported by an NSERC Discovery Grant} \quad Mark Penney\footnote{Current address: Mathematical Institute, University of Oxford,
Andrew Wiles Building,
Radcliffe Observatory Quarter,
Woodstock Road,
Oxford, OX2 6GG;  mark.penney@maths.ox.ac.uk}}

\date{March 10, 2015}

\maketitle

\abstract{We investigate the process of quantum measurements on scattered probes. Before scattering, the probes are independent, but they become entangled afterwards, due to the interaction with the scatterer. The collection of measurement results (the history) is a stochastic process of dependent random variables. We link the asymptotic properties of this process to spectral characteristics of the dynamics. We show that the process has decaying time correlations and that a zero-one law holds. We deduce that if the incoming probes are not sharply localized with respect to the spectrum of the measurement operator, then the process does not converge. Nevertheless, the scattering modifies the measurement outcome frequencies, which are shown to be the average of the measurement projection operator, evolved for one interaction period, in an asymptotic state.  We illustrate the results on a truncated Jaynes-Cummings model.}


\section{Introduction and main results}

We consider a scattering experiment in which a beam of probes is directed at a scatterer. The probes are sent to interact sequentially, one by one. Before the scattering process, they are identical and independent. The interaction of each probe with the system is governed by a fixed interaction time $\tau>0$ and a fixed interaction operator $V$.  After interacting with the scatterer, a quantum measurement is performed on each ``outcoming'' probe. The result of the measurement of the $n$-th probe is a random variable, denoted $X_n$. The stochastic process $\{X_n\}_{n\geq 1}$ is the measurement history. Due to entanglement of the probes with the scatterer, the $X_n$ are not independent random variables. We analyze asymptotic properties of this process.

A concrete physical setup is given by atoms (being the probes) shot through a cavity containing an electromagnetic field, the modes which interact with the atoms forming the scatterer. We assume that the incoming probe states are stationary with respect to their isolated dynamics.

We study systems with only finitely many degrees of freedom involved in the scattering process. This means that the Hilbert spaces of pure states both of the system and each probe is finite-dimensional. The measurement of a probe is a von Neumann, or projective, measurement associated to a self-adjoint probe measurement observable $M$. The eigenvalues $m$ of $M$ are the possible measurement outcomes. Due to finite dimensionality, the random variables $X_n$ have finite range.

The present work can be viewed as the continuation of recently developed techniques for the mathematical analysis of repeated interaction quantum systems  \cite{BJM1,BJM2,BJM3,BJM4}. In these references, asymptotic properties of the scatterer have been investigated, without considering the fate of the outcoming probes, and without quantum measurements. While the setup of our present work is similar to the one in the given references, our focus here is on the measurement outcomes process. We show that generically, this process {\em does not converge}. We describe the fluctuations on the measurement history, provoked by the scattering process, by analyzing the measurement frequencies. A more detailed comparison to related works is given at the end of this section.

As explained in the references above, in absence of quantum measurements on probes, and under a generic ergodicity assumption, one shows that the scatterer approaches a so-called {\it repeated interaction asymptotic state} after many interactions. We keep this assumption in the present work.

\begin{itemize}
\item[{\bf (A)}] Assume that if no measurement is performed ($M=\bbbone$), then, under the repeated interaction with the probes, the scatterer approaches a final state. The convergence is exponentially quick in time.
\end{itemize}

The precise mathematical formulation of this assumption is given in Section \ref{sectanalproba}, see before \fer{mm35}. It is a condition on the spectrum of a reduced dynamics operator, and necessitates the introduction of some technicalities which we want to avoid in this introduction. Condition (A) is generically satisfied, and is not hard to be verified explicitly, and one even calculates the rate of convergence for concrete models (see the above references).

\medskip

We now explain our main results. 
Denote by $\sigma(X_r,\ldots,X_s)$ the {\it sigma-algebra generated by the random variables $X_r,\ldots,X_s$}, $1\leq r\leq s\leq \infty$. We denote by $P$ the probability measure associated with the process $\{X_n\}_{n\geq 1}$. 
\begin{thm}[Decay of correlations]
\label{corrdeclemma'}
Suppose that Condition {\rm (A)} holds. There are constants $c$, $\gamma'>0$, such that for $1\leq k\leq l < m\leq n<\infty$, $A\in\sigma(X_k,\ldots X_l)$ and $B\in\sigma(X_m,\ldots,X_n)$, we have 
\begin{equation}
\left| P(A\cap B)-P(A)P(B)\right| \leq c P(A) \ \e^{-\gamma'(m-l)}.
\label{010'}
\end{equation}
\end{thm}
 We give a proof of the theorem in Section \ref{proof1+5}. Intuitively, the system starts relaxation to its asymptotic state during the time $m-l$ between two consecutive measurements, and hence erases correlations between the two measurements. The rate $\gamma'$ in \fer{010'} is linked to the convergence rate in Assumption (A), see Section \ref{sectanalproba}.

\medskip
The {\it tail sigma-algebra} is defined by ${\cal T}= \cap_{n\geq 1} \sigma(X_n, X_{n+1},\ldots)$. Decaying correlations imply the following {\it zero-one law}.
\begin{trace}[Zero-one law]
\label{thmtailevent'} Assume that Condition {\rm (A)} holds. 
Any tail event $A\in{\cal T}$ satisfies $P(A)=0$ or $P(A)=1$.
\end{trace}
In textbooks, the Kolmogorov zero-one law is usually presented for {\it independent} random variables \cite{B}. However, an adaptation of the proof yields the result for random variables with decaying correlations, see \cite{A} (and also Section \ref{sectanalproba}). The tail sigma-algebra captures convergence properties. For instance, given any outcome $m\in{\rm spec}(M)$, the set $\{\lim_nX_n=m\}$ is a tail event, hence, according to Corollary \ref{thmtailevent'}, it has probability zero or one.

\medskip

We now explain why fluctuations in the process persist generically, for all times. Let $\omega_{\rm in}$ be the state of the incoming probes, denote by $E_S$ the spectral projection of the measurement operator $M$ associated to $S\subset{\rm spec}(M)$ and denote $E_m=E_{\{m\}}$ for $m\in{\rm spec}(M)$. In absence of interaction ($V=0$ or $\tau=0$), the $X_j$ are independent random variables. We show in Proposition \ref{propx2} that the dependence generated by the interaction with the scatterer is small for small interactions, {\it uniformly in time}. Therefore, since $P(X_n=m)=\omega_{\rm in}(E_m)+ O(\|V\|)$, we have $P(X_{n+1}=m, X_n=m)=P(X_{n+1}=m)P(X_n=m) +O(\|V\|)$, and consequently,
$$
P(X_{n+1}=X_n) = \sum_{m\in{\rm spec}(M)}\omega_{\rm in}^2(E_m) +O(\|V\|).
$$
The numbers $\omega_{\rm in}(E_m)$ are probabilities. Thus, $\sum_m\omega_{\rm in}^2(E_m)=1$ if and only if for a single $m_0$ we have $\omega_{\rm in}(E_{m_0})=1$ while for all other $m$, $\omega_{\rm in}(E_m)=0$. This means that $P(X_{n+1}=X_n)<1$ for small $V$, whenever there are several $m$ with $\omega_{\rm in}(E_m)>0$. Together with the zero-one law, this implies that {\em $P(X_n {\rm \ converges})=0$ whenever the incoming state is not localized in a single subspace of $M$ (and $V$ is small enough).} If $m$ is a simple eigenvalue of $M$ with associated eigenvector $\psi_m$, then $\omega_{\rm in}(E_m)=1$ is equivalent to $\omega_{\rm in}(\cdot)=\scalprod{\psi_m}{\cdot \,\psi_m}$. Statistical fluctuations in the incoming probes (mixture of states localized w.r.t. measurement values) thus get transferred to outcoming probes, even in the limit of large times. The following is a more general statement of this fact.
\begin{thm}
\label{propx1} Assume Condition {\rm (A)} holds. There is a constant $C$ s.t., for any $S\subset {\rm spec}(M)$ with $\omega_{\rm in}(E_S)\neq 1$, if $\|V\|\leq C(1-\omega_{\rm in}(E_S))$, then 
$$
P(X_n\in S \rm{\ eventually})=0.
$$
\end{thm}
The result on non-convergence of $X_n$ explained before Theorem \ref{propx1} is a special case of Theorem \ref{propx1}, when $S=\{m\}$, $m\in{\rm spec}(M)$. We mention that our analysis also gives a condition under which $P(X_n\in S \rm{\ eventually})=1$, see Lemma \ref{lemma100}.

\medskip
The process $X_n$ carries information about the scattering process, encoded in the relative occurrence of a particular measurement outcome. We define the {\it frequency} of $m\in{\rm spec}(M)$ by 
\begin{equation*}
f_m = \lim_{n\rightarrow\infty} \frac 1n\{\mbox{number of $k\in\{1,\ldots,n\}$ s.t. $X_k=m$}\}.
\end{equation*}
$f_m$ is {\em a priori} a random variable and the existence (and type) of limit has to be clarified.

\begin{thm}[Frequencies]
\label{theorem2}
Assume condition {\rm (A)} holds, and denote the final scatterer state under the evolution without measurements by $\omega_+$. Let $H$ be the interacting Hamiltonian of a single probe with the scatterer. Then the frequencies $f_m$ exist (as almost everywhere limits) and are deterministic (not random), given by 
$$
f_m =\omega_+\otimes\omega_{\rm in}\big(\e^{\i \tau H} E_m \e^{-\i\tau H}\big).
$$
\end{thm}

{\em Remark.\ } The proof of Theorem \ref{theorem2}, given in Section \ref{proofthm2}, can be readily generalized to yield the following result: For any $m\ge1$, $S_1,\ldots, S_m\subset\spec$, 
\begin{eqnarray*}
\lefteqn{
\lim_{n\rightarrow\infty}\frac1n \big\{ \mbox{ number of $j\le n$ s.t. $X_j\in S_1,\ldots, X_{j+m}\in S_m$}\big\}}\\
&=&\omega_+\otimes\omega_{\rm in}\cdots\otimes\omega_{\rm in} \big( \e^{\i\tau H_1}\cdots\e^{\i\tau H_m} E_{S_1}\cdots E_{S_m} \e^{-\i\tau H_1}\cdots\e^{-\i\tau H_m}\big).
\end{eqnarray*}
Here, $H_j$ is the Hamiltonian describing the free motion of $\s$ and $m$ probes $\p$, plus the interaction of $\s$ with the $j$-th probe.

The next result describes the process $\{\overline{\! X}_{\! n}\}$ of the empirical average
$$
\overline{\! X}_{\! n} = \frac1n(X_1+\cdots +X_n).
$$
\begin{thm}[Mean]
\label{prop01'}
Assume condition (A) holds and adopt the notation of Theorem \ref{theorem2}. We have a law of large numbers, 
$$
\lim_{n\rightarrow\infty} \overline{\! X}_{\! n}=\mu_\infty:=\omega_+\otimes\omega_{\rm in}\big(\e^{\i\tau H}M\e^{-\i\tau H}\big).
$$
The limit is in the almost everywhere sense. Note that $\mu_\infty =\sum_m mf_m$.
\end{thm}

\medskip
{\bf Relation to other work.\ } The literature on repeated interaction systems we are aware of can be classified into two categories. In a first one, effective evolution equations are derived by taking continuous interaction limits \cite{a1,a2,ay1,ay2,P,PP,D} and in a second category, the dynamics is left discrete, but the time-asymptotics is investigated \cite{BJM1,BJM2,BJM3,BJM4,BB,BBB,DH,NTY}. To our knowledge, repeated interaction systems have been first proposed in \cite{Kuemmerer,KM0} and in \cite{a1,a2} as approximations for system-environment type models, where it is proven that the discrete evolution converges to that of a quantum Langevin equation, in the limit of ever shorter system-probe interaction times. The study of the time-asymptotics was initiated in \cite{BJM1} (and continued in \cite{BJM2,BJM3,BJM4}). It is shown there that the (reduced) system state converges to a final state and this state's thermodynamic properties were analyzed rigorously.

The mathematical formalism in \cite{BJM1,BJM2,BJM3,BJM4} is based on a spectral approach to the time-asymptotics of open quantum systems, expressed in the language of operator algebraic quantum theory. The present work extends this formalism to the setting of repeated probe quantum measurements. A review of  the above-cited works, as well as an announcement of some of the results of the present work, can be found in \cite{BJM5}.

In \cite{KM} a result similar to Corollary \ref{thmtailevent'}, is shown. There, the authors discuss ergodic properties of quantum counting processes, described by unravellings of a Lindblad generator. The analysis is based on the assumption that the dynamics, without taking into account the counting, is {\em ergodic} (i.e., converges in the mean to an equilibrium state). On the other hand, our system is {\em mixing} due to Condition A (i.e., it converges `pointwise').

In \cite{BB,BBB}, an energy conserving model is considered, where the dynamics is assumed to be such that there exists a basis of special stationary states (`pointer states'). Then the system converges to one of these states, determined by the measurement outcomes. Many physical systems do not fall into this class. Whenever there is exchange of energy between probes and the system, there is only one invariant state (under generic conditions). An important example of an energy-exchange system is  the `one atom maser', where atoms (probes) interact with modes of the electromagnetic field in a cavity (system) by exciting the field modes, leading to subsequent photon emission \cite{MWM}. A famous mathematical model describing this situation is the Jaynes-Cummings model, a simplification of which we discuss in Section \ref{example}.

\section{Quantum dynamical system setup}
\label{qdssetup}

\subsection{Formalism}
\label{subsectionformalism}

The general formalism of quantum dynamical systems is presented in \cite{AJP,BR} (see also \cite{BJM1}). 
Both $\s$ and $\p$ are described as quantum ($W^*$) dynamical systems in standard form. The states of such a system are given by unit vectors in a Hilbert space $\ch$, observables form a von Neumann algebra $\cm\subset{\cal B}(\h)$ and the dynamics is given by a group of *automorphisms $\alpha^t$ on $\cm$, $t\in\mathbb R$. There is a distinguished (reference) vector $\psi\in\ch$ which is cyclic and separating for $\cm$, and such that the dynamics is represented as $\alpha^t(A)=\ee^{\i tL} A\ee^{-\i tL}$, for $A\in\cm$, and where $L$ is a selfadjoint operator on $\ch$ satisfying $L\psi=0$. This operator is called the {\it standard Liouville operator}.

Accordingly, the system $\s$ is determined by a Hilbert space $\ch_\s$, a von Neumann algebra $\cm_\s$, a cyclic and separating vector $\psi_\s$ and a dynamics $\alpha^t_\s = \ee^{\i tL_\s}\cdot \ee^{-\i tL_\s}$, with $L_\s\psi_\s=0$. A single probe is described by the same ingredients (with index $\s$ replaced by $\p$). We assume throughout the paper that $\dim\h_\s<\infty$, $\dim\ch_\p<\infty$. The Hilbert space of the chain of all probes is the tensor product $\ch_\c=\otimes_{n\geq 1}\ch_\p$, stabilized on the reference vector $\psi_\p\in\ch_\p$. The von Neumann algebra of observables of $\c$ is $\cm_\c=\otimes_{n\geq 1}\cm_\p$ and its dynamics is $\alpha^t_\c=\otimes_{n\geq 1}\alpha^t_\p$.

The full system is described by the Hilbert space 
\begin{equation}
\ch= \ch_\s\otimes\ch_\c
\label{mm01}
\end{equation}
on which acts the von Neumann algebra of observables
\begin{equation}
\cm= \cm_\s\otimes\cm_\c.
\label{mm02}
\end{equation}
The non-interacting Liouville operator is given by
\begin{equation}
\widetilde L_0 = L_\s+\sum_{n\geq 1} L_{n,\p},
\label{mm03}
\end{equation}
where $L_{n,\p}$ is the operator acting trivially on all factors of $\ch$ except on the $n$-th factor of $\ch_\c$, on which it acts as $L_\p$. The interaction between $\s$ and the $n$-th probe lasts for a duration of $\tau>0$ and is determined by an operator $V_n$, acting trivially on all factors of $\ch$ except $\ch_\s$ and the $n$-th one in $\ch_\c$, where it acts as a fixed selfadjoint operator
\begin{equation}
V=V^*\in\cm_\s\otimes\cm_\p.
\label{mm04}
\end{equation}
Let $\Psi_0\in\ch$ be an initial state of the full system. Then the state of the system at time-step $n$ is given by the vector
\begin{equation}
\Psi_n = U_n\cdots U_2 U_1\Psi_0,
\label{mm05}
\end{equation}
where 
\begin{equation}
U_n = \ee^{-\i\tau ({\widetilde L}_0+V_n)}
\label{mm06}
\end{equation}
is the unitary generating the one-step time evolution at instant $n$.

We consider initial states of the form 
\begin{equation}
\Psi_0 = \Psi_\s\otimes_{n\geq 1} B\psi_\p,
\label{mm07}
\end{equation}
where $B$ is an operator in the commutant von Neumann algebra 
$$
\cm'_{\cal C} = \{A\in{\cal B}(\ch_{\cal P})\ :\ AX=XA \ \forall X\in\cm_{\cal C} \},
$$
such that $\|B\psi_\p\|=1$ and $L_\p B=BL_\p$ (normalized, invariant state). 
Since $\psi_\p$ is cyclic for $\cm'_{\cal P}$ ($\Leftrightarrow$ separating for $\cm_{\cal P}$, where $\cm'_{\cal P}$ the commutant of $\cm_\p$) and since $\dim \ch_\p$ is finite, every $\psi\in\ch_\p$ is exactly represented as $\psi=B\psi_\p$ for a unique $B\in\cm'_{\cal P}$. 

{\bf Remarks.\ } 1. Since both $\cal H_S$ and $\cal H_P$ are finite-dimensional, one can work in a less general framework and analyze the measurement process based on a density matrix description of the system and the probes, as opposed to work in the (GNS) Hilbert space representation.  Such an analysis can be carried out even for some systems having infinitely many discrete energy levels, see  \cite{BrPi}. Nevertheless, we adhere to the present setup. It will come in handy (and necessary) when one considers models where the system is large (e.g. some membrane or screen) subjected to an incoming beam of scattering probes.

2. We have presented in this section the chain Hilbert space as an {\em infinite tensor product} of single-probe spaces, and similarly for the evolution and reference state. All quantities we are examining involve the whole system up to arbitrary but {\em finite} times, involving only {\em finitely} many probes. Consequently, we could define our Hilbert space to be time-dependent, having $n$ probe factors only, but with $n$ arbitrarily large. The expressions for the probabilities of the measurement process, or the reduced state of the scatterer, or the exited probes, would not change. Whether we take an infinite tensor product or a finite one, with arbitrary factors, does not influence the physical properties we describe. Nevertheless, mathematically, there is a slight difference between these two cases, see e.g. \cite{BJM1,BJM2,BJM3,BJM4,BJM5}. As an example, strictly speaking, the vector $\Psi_0$ given in \eqref{mm07} does not belong to the Hilbert space ${\cal H}$ unless $B=\bbbone$ (because ${\cal H}_\c$ is ``stabilized'' along $\psi_\p$). What we mean is that, at arbitrary time $n$, the vector $\Psi_0$ has the form $\Psi_\s\otimes B\psi_\p\otimes B\psi_\p\otimes\cdots\otimes B\psi_\p$ ($n$ probe factors) and as such belongs to the Hilbert space ${\cal H}_\s\otimes{\cal H}_\p\otimes{\cal H}_\p\otimes\cdots\otimes{\cal H}_\p$. 

\medskip

Let $J$ and $\Delta$ denote the modular conjugation and the modular operator associated to the cyclic and separating vector $\psi_\s\otimes\psi_\p$ for the von Neumann algebra $\cm_\s\otimes\cm_\p$. By the Tomita-Takesaki theorem \cite{B}, we know that $\Delta^{\i t}(\cm_\s\otimes\cm_\p)\Delta^{-\i t}=\cm_\s\otimes\cm_\p$ for all $t\in\mathbb R$ ($\Delta^{-\i t}$ is a group of unitaries), and that $J(\cm_\s\otimes\cm_\p) J = \cm_\s'\otimes\cm_\p'$. Consequently, $J\Delta^{\i t} V\Delta^{-\i t}J\in \cm_\s'\otimes\cm_\p'$, where $V$ is the interaction operator defined in \fer{mm04}. In the finite-dimensional case as considered here, an easy analyticity argument shows that the last relation stays valid for any $t\in{\mathbb C}$. In particular, 
\begin{equation}
J\Delta^{1/2} V\Delta^{-1/2}J\in \cm_\s'\otimes\cm_\p'.
\label{mm33}
\end{equation}

It will be convenient to represent the joint dynamics of the system $\s$ and the probe $\p$ interacting at the given moment with $\s$ by the following operator acting on $\ch_\s\otimes\ch_\p$ (see Subsection \ref{andyn}),
\begin{equation}
K = L_\s+L_\p+\lambda V - \lambda J\Delta^{1/2}V\Delta^{-1/2}J.
\label{mm21}
\end{equation}
Here, $\lambda\in\mathbb R$ is a coupling constant. Due to property \fer{mm33} we have 
\begin{equation}
\ee^{\i t K} A \ee^{-\i t K} = \ee^{\i t[L_\s+L_\p+\lambda V]} A \ee^{-\i t [L_\s+L_\p+\lambda V]}
\label{f11}
\end{equation}
for all $A\in \cm_\s\otimes\cm_\p$ and all $t\in\mathbb R$, as is not hard to see for instance by using the Trotter product formula. The term $- \lambda J\Delta^{1/2}V\Delta^{-1/2}J$ in \fer{mm21} is introduced in order to have the property
\begin{equation}
K \psi_\s\otimes\psi_\p =0.
\label{mm34}
\end{equation}
The latter relation follows from $(L_\s+L_\p)\psi_\s\otimes\psi_\p =0$ and the fact that $J\Delta^{1/2}V\Delta^{-1/2}J\psi_\s\otimes\psi_\p =V\psi_\s\otimes\psi_\p$ (which in turn is implied by $\Delta^{-1/2}J=J\Delta^{1/2}$ and $J\Delta^{1/2} A
\psi_\s\otimes\psi_\p = A^*\psi_\s\otimes\psi_\p$ for all $A\in\cm_\s\otimes\cm_\p$). The operator $K$ has been used in \cite{JP} for the study of non-equilibrium open quantum systems and in the setting of repeated interaction open systems in \cite{BJM1,BJM2,BJM3,BJM4}.

\subsection{Multitime measurement process}

In this subsection, we describe the process of multitime measurement of the outcoming probes. We refer to \cite{NC,BP,GZ,C} for a detailed introduction to quantum measurement theory, outlining here only a few particularities pertaining to repeated interaction systems. Consider a quantum system described by a density matrix $\rho$ and a self-adjoint ``measurement operator'' $M$ having eigenvalues ${\rm spec}(M)=\{m_1,\ldots,m_\mu\}$ with corresponding eigenprojections $P_1,\ldots,P_\mu$. When the system undergoes interaction with the measurement apparatus and we know that the measurement result is one of the eigenvalues from a set $S\subset {\rm spec}(M)$, then the state after measurement is given by
\begin{equation}
\label{r0}
\rho' = \sum_{\{j :  m_j\in S\}} P_j\rho P_j.
\end{equation}
This corresponds to a non-selective measurement \cite{GZ}. In the repeated interaction setting, the following experiment is performed: the entire system evolves according to $U_1$ (interaction between the first probe and the system) and then a measurement of the observable $M$ is made on the outcoming probe, yielding a value in $S_1\subset {\rm spec}(M)$, and then the system evolves according to $U_2$ (interaction with second probe) and after this evolution a measurement of $M$ is made on the outcoming probe and yields a result lying in $S_2\subset {\rm spec}(M)$, and this procedure is repeated $n$ times. Let $\widetilde\U_j=\e^{-\i \tau \widetilde H_j}$, where $\widetilde H_j=H_\s+H_{1,\p}+\cdots +H_{n,\p}+V_j$ and  $V_j$ is the interaction operator, acting on the system and the $j$th probe only. The expectation of a system observable $O_\s$ after $n$ probe measurements have been performed, knowing that measurement $j$ has yielded a result in the set $S_j\subset {\rm spec}(M)$, is given by 
\begin{eqnarray}
\label{r1}
\lefteqn{
\omega_{S_1,\ldots,S_n}(O_\s)}\\
&& = \frac{1}{P(S_1,\ldots, S_n)}\sum_{\{j_n : m_{j_n}\in S_n\}}\cdots\sum_{\{j_1 : m_{j_1}\in S_1\}} {\rm Tr}\big( P_{j_n}\widetilde \U_n \cdots P_{j_1} \widetilde \U_1\rho\, \widetilde \U_1^*P_{j_1}\cdots \widetilde \U_n^* P_{j_n}  O_\s \big). \nonumber
\end{eqnarray}
Here, $P(S_1,\ldots,S_n)$ is the probability that the measurement outcomes at time $j$ lies in the set $S_j$, for $j=1,\ldots,n$, and is determined by $\omega_{S_1,\ldots,S_n}(\bbbone ) = 1$. 
Using the cyclicity of the trace and the structure of the Hamiltonians $\widetilde H_j$, $j=1,\ldots,n$, one readily sees that 
\begin{eqnarray}
\label{r3}
\lefteqn{
\omega_{S_1,\ldots,S_n}(O_\s)}\\
&& = \frac{1}{P(S_1,\ldots, S_n)} {\rm Tr}\Big( \rho\,  (\U_n^+)^* \,  \U_1^* E_{S_1}\cdots \U_n^*E_{S_n}  O_\s  E_{S_n}\U_n\cdots E_{S_1} \U_1 \,\U^+_n\Big), \nonumber
\end{eqnarray}
where $E_S=\sum_{\{j : m_j\in S\}}P_j$, $\U^+_n=\e^{-\i\tau\sum_{j=2}^n(j-1)H_{j,\p}}$ and $\U_j = \e^{-\i\tau (H_\s+H_{j,\p}+V_j)}$. It follows directly from \eqref{r3} that if we do not know about the outcome of any of the measurements, which corresponds to $S_j={\rm spec}(M)$ and $E_{S_j}=\bbbone$, for all $j=1,\ldots,n$, then the dynamics of the system is the same as if no interaction with the measurement apparatus took place. This phenomenon is particular to the setup of repeated interactions and is not true for general quantum systems. Indeed, relation \eqref{r0} with the sum extended over all $j$ does not yield $\rho'=\rho$ in general. 

Relation \eqref{r3} implies that 
\begin{equation}
\label{r4}
P(S_1,\ldots,S_n) = {\rm Tr}\Big( \rho\,  (\U_n^+)^* \, \widetilde \U_1^* E_{S_1}\cdots \widetilde\U_n^*E_{S_n}\widetilde\U_n\cdots E_{S_1}\widetilde \U_1 \,\U^+_n\Big).
\end{equation}
The stochastic process associated to the measurements is constructed as follows. Let 
\begin{equation*}
\Omega=\Sigma^{\mathbb N}=\{ \omega=(\omega_1,\omega_2,\ldots)\ : \ \omega_j\in{\rm spec}(M)\}
\end{equation*}
and let $\cal F$ be the $\sigma$-algebra of subsets of $\Omega$ generated by all cylinder sets of the form
$$
\{\omega\in\Omega\ :\ \omega_1\in S_1, \ldots, \omega_n\in S_n, \ n\in{\mathbb N}, \ S_j\subset {\rm spec}(M)\}.
$$
On $(\Omega,{\cal F})$ we define the random variables $X_n:\Omega\rightarrow {\rm spec}(M)$ by $X_n(\omega)=\omega_n$, for $n=1,2,\ldots$ The random variable $X_n$ represents the outcome of the measurement at time-step $n$. The finite-dimensional distribution of the process $\{X_n\}_{n\geq 1}$ is given by 
\begin{equation}
P(X_1\in S_1,\ldots, X_n\in S_n) = P(S_1,\ldots,S_n),
\label{mm3}
\end{equation}
for any $n\in \mathbb N$, any subsets $S_1,\ldots,S_n$ of ${\rm spec}(M)$ and where the right hand side is defined in \fer{r4}. $P$ extends uniquely to a probability measure on $(\Omega,{\cal F})$ by the Kolmogorov extension theorem.

\subsection{Representation of joint probabilities}
\label{andyn}

In the (GNS) Hilbert space setting, the multi-time measurement process introduced in the previous paragraph is formulated as follows. Let $M\in\cm_\p$ be a selfadjoint ``measurement'' operator on $\h_\p$ with spectrum  ${\rm spec}(M)=\{m_1,\ldots, m_\mu\}$, 
where $1\leq \mu\leq \dim\h_{\cal P}$ (distinct eigenvalues). Let $S$ be any subset of ${\rm spec}(M)$ and denote by $E_S$ the spectral projection of $M$ associated to $S$. Suppose that the entire system is in a state $\Psi_0\in\ch$ initially (see also Remark 2 in the previous paragraph). The state of the system after $n$ measurements, viewed as a state of $\cm_\s$, is given by (see \eqref{r3}) $A\mapsto \av{A}_n = \scalprod{\Psi_n}{A\Psi_n}$, where
\begin{equation}
\label{r5}
\Psi_n = \|\widetilde\Psi_n\|^{-1}\widetilde\Psi_n\quad \mbox{and} \quad \widetilde\Psi_n =  E_{S_n} \ee^{-\i\tau L_n}\cdots E_{S_1} \ee^{-i\tau L_1} U_n^+ \Psi_0.
\end{equation}
Here, 
\begin{equation}
L_j = L_\s+L_{j,\p}+V_j
\label{mm11}
\end{equation}
acts non-trivially only on the Hilbert space of $\s$ and the $j$-th probe Hilbert space, and  we introduced the unitary
\begin{equation}
U_n^+ = \exp\left[-\i\tau\sum_{j=2}^n (j-1)L_{j,\p}\right].
\label{mm10}
\end{equation}
Recall that $\Psi_0=(\bbbone_\s\otimes B\otimes B\cdots)\Psi_{\rm ref}$, with $\Psi_{\rm ref}=\psi_\s\otimes\psi_\p\otimes\psi_\p\cdots$. The operator $C_j=\ee^{-\i\tau(j-1)L_{j,\p}} B \ee^{\i\tau(j-1)L_{j,\p}}$ belongs to the commutant $\cm_\p'$, since $B$ does, and since the dynamics generated by $L_{j,\p}$ leaves $\cm_\p'$ invariant. We obtain 
\begin{eqnarray}
\lefteqn{\scalprod{\widetilde \Psi_n}{A\widetilde\Psi_n}}\nonumber\\
&=&
\scalprod{C_1 \cdots C_n\Psi_{\rm ref}}{\ee^{\i\tau L_1}E_{S_1}\cdots \ee^{\i\tau L_n}E_{S_n} A E_{S_n} \ee^{-\i\tau L_n}\cdots E_{S_1} \ee^{-\i\tau L_1} C_1 \cdots C_n\Psi_{\rm ref}}\nonumber\\
&=& \scalprod{\Psi_{\rm ref}}{C_1^*C_1\cdots C_n^*C_n \ee^{\i\tau K_1}E_{S_1} \cdots \ee^{\i\tau K_n}E_{S_n} A \Psi_{\rm ref}} \nonumber\\
&=& \scalprod{\Psi_{\rm ref}}{[P_1C_1^*C_1\ee^{\i\tau K_1}E_{S_1}P_1]\cdots [P_nC_n^*C_n \ee^{\i\tau K_n}E_{S_n}P_n] A \Psi_{\rm ref}}, 
\label{mm13}
\end{eqnarray}
where $P_j$ is the projection acting trivially on all factors of $\ch$ except on the $j$-th $\ch_\p$, where it acts as the rank-one orthogonal projection onto $\psi_\p$. We define the operator $T_j = P_jC^*_jC_j \ee^{\i\tau K_j} E_{S_j}P_j$, which, under the hypothesis that $\ee^{\i s L_\p}B\ee^{-\i sL_\p}=B$, becomes $T_j = P_jB^*B \ee^{\i\tau K_j} E_{S_j}P_j$. We identify $T_j$ as an operator on the Hilbert space $\ch_\s$, and as such, write
\begin{equation}
T_S = P B^*B \ee^{\i\tau K}E_{S}P,
\label{mm14}
\end{equation}
where $P$ the orthogonal projection onto $\psi_\p\otimes \ch_\c$, $S\subset\spec$ determines the measurement performed at the given time-step, and where $K$ is given in \fer{mm21}. We write simply $T$ for $T_{{\rm spec}(M)}$. Remark that since $E_{S_j}\in\cm_\p$ and $\psi_\p$ is separating for $\cm_\p$, we have $E_{S_j}P_j \neq 0$ for all $j$. With this definition, we arrive at
\begin{equation}
\scalprod{\widetilde \Psi_n}{A\widetilde\Psi_n} = \scalprod{\psi_\s}{T_{S_1} T_{S_2}\cdots T_{S_n} A\psi_\s}.
\label{mm15}
\end{equation}
In particular, measurement probability can be expressed as
\begin{equation}
P(S_1,\ldots, S_n) = \scalprod{\psi_\s}{T_{S_1}\cdots T_{S_n}\psi_\s}.
\label{mm16}
\end{equation}

\subsection{Analysis of joint probabilities}

\label{sectanalproba}

\begin{lem}
\label{lemma1}
The spectrum of $T_S$, \fer{mm14}, lies in the closed unit disk centered at the origin of the complex plane. For $S={\rm spec}(M)$, i.e., $E_S=\bbbone$, we have in addition $T\psi_\s=\psi_\s$.
\end{lem}

We consider the probability $P(X_n\in S \mbox{\ eventually})$, for $S\subset{\rm spec}(M)$. This quantity can be expressed using the Riesz spectral projections $\Pi$ and $\Pi_S$ of the operators $T$ and $T_S$ associated to the eigenvalue $1$. They are defined by
\begin{equation}
\Pi_S = \frac{1}{2\pi \i}\oint (z-T_S)^{-1} {\rm d}z,\qquad \Pi=\Pi_{{\rm spec}(M)},
\label{projections}
\end{equation}
where the integral is over a simple closed contour in the complex plane encircling no spectrum of $T_S$ except the point $1$. If $1$ is not an eigenvalue then $\Pi_S=0$. For the next result, we recall the following definition,
$$
\{ X_n\in S \mbox{\  eventually\ }\}=\{ \omega |  \mbox{ there exists a $k$ s.t. $X_n(\omega)\in S$ for all $n\geq k$}\}.
$$

\begin{lem}
\label{lem3}
We have $P(X_n\in S \mbox{\ eventually}) = \scalprod{\psi_\s}{\Pi\,\Pi_S\,\psi_\s}$.
\end{lem}

{\it Proof.\ } The set $\{X_n\in S \mbox{\ eventually}\}$ is the increasing union of $\{X_n\in S \ \forall n\geq k\}$, so $P(X_n\in S \mbox{\ eventually}) =\lim_{k\rightarrow\infty} P(X_n\in S\ \forall n\geq k)$. Next, $\{X_n\in S\ \forall n\geq k\}$ is the intersection of the decreasing sequence $\{X_n\in S,\ n=k,\ldots,k+l\}$, so 
\begin{equation}
P(X_n\in S \mbox{\ eventually}) = \lim_{k\rightarrow\infty}\lim_{l\rightarrow\infty} P(X_n\in S,\ n=k,\ldots,k+l). \label{dl1}
\end{equation}
We have $P(X_n\in S,\ n=k,\ldots,k+l)=\scalprod{\psi_\s}{ T^{k-1} T_S^{l+1}\psi_\s}$. Since for each $k$ fixed, the limit of $\scalprod{\psi_\s}{ T^{k-1} T_S^{l+1}\psi_\s}$ as $l\rightarrow\infty$ exists (it is the probability $P(X_n\in S \ \forall n\geq k)$), we have 
$$
\lim_{l\rightarrow\infty}\scalprod{\psi_\s}{ T^{k-1} T_S^{l+1}\psi_\s} = \lim_{L\rightarrow\infty} \frac 1L \sum_{l=1}^L \scalprod{\psi_\s}{ T^{k-1} T_S^{l+1}\psi_\s} = \scalprod{\psi_\s}{ T^{k-1} \widetilde\Pi_S\psi_\s},
$$
where $\widetilde \Pi_S$ is the ergodic projection of $T_S$ associated to the eigenvalue $1$. ($\widetilde\Pi_S=0$ if 1 is not an eigenvalue of $T_S$.) Arguing in the same way for the limit $k\rightarrow\infty$, we obtain $P(X_n\in S \mbox{\ eventually}) = \textstyle\scalprod{\psi_\s}{ \widetilde\Pi\, \widetilde\Pi_S\, \psi_\s}$. 
Invoking Lemma \ref{lemma10} we replace the ergodic projections with the Riesz projections. \hfill $\blacksquare$

\medskip

The probability is given entirely by information on the spectrum of $T$ and $T_S$ at the point 1 (Riesz projections). The result holds even if $T$ or $T_S$ have spectrum on the unit circle other than possibly at 1. 

Finer information about the asymptotic dynamics depends on all the spectrum of $T$ on the unit circle. We make the following ergodicity assumption (compare with Lemma \ref{lemma1}).

\medskip
\noindent
{\bf Condition A.} {\it The point $z=1$ is the unique eigenvalue of $T$ with $|z|=1$, and this eigenvalue is simple (with eigenvector $\psi_\s$). We define the gap by
\begin{equation}
\gamma =1- \sup\{ |z|\ :\ z\in{\rm spec}(T), z\neq 1\}.
\label{mm35}
\end{equation}}
{} For $\lambda=0$ the operator $T$ is $\ee^{\i \tau L_\s}$ and has spectrum on the unit circle, with degenerate eigenvalue 1. Assumption A is verified in practice typically by perturbation theory ($\lambda$ small, nonzero). It is sometimes called a ``Fermi golden rule condition''.
In this setting, condition A implies $\gamma>0$ for small nonzero $\lambda$. Condition A implies the dynamical behaviour of the assumption (A) stated before Theorem \ref{corrdeclemma'}.

The random variables $X_n$ are not independent but their correlations decay.

\begin{thm}[Decay of correlations]
\label{corrdeclemma}
Suppose Condition A holds. For any $\epsilon>0$ there is a constant $C_\epsilon$ such that for $1\leq k\leq l < m\leq n<\infty$, $A\in\sigma(X_k,\ldots X_l)$ and $B\in\sigma(X_m,\ldots,X_n)$, we have 
\begin{equation}
\left| P(A\cap B)-P(A)P(B)\right| \leq C_\epsilon P(A) \ \e^{-(m-l)[\ln(\frac{1}{1-\gamma})-\epsilon]}.
\label{010}
\end{equation}
\end{thm}

We give a proof of Theorem \ref{corrdeclemma} in Section \ref{proof1+5}. 
For fixed $\epsilon<\ln(\frac{1}{1-\gamma})$, define the function ${\cal C}:{\mathbb N} \rightarrow {\mathbb R}_+$ by
\begin{equation}
{\cal C}(d) = C_\epsilon \e^{-d[\ln(\frac{1}{1-\gamma})-\epsilon]}. 
\label{015}
\end{equation}
Theorem \ref{corrdeclemma} implies that the random variables $X_n$ have decaying correlations in the following sense: for all $1\leq k\leq l < m\leq n$, all $A\in\sigma(X_k,\ldots X_l)$ and all $B\in\sigma(X_m,\ldots,X_n)$, we have 
\begin{equation}
\left| P(A\cap B)-P(A)P(B)\right| \leq {\cal C}(m-l),
\label{016}
\end{equation}
where the function ${\cal C}$ is independent of $A,B,k,l,m,n$ and satisfies ${\cal C}(d)\rightarrow 0$ as $d\rightarrow\infty$.

Let $X_k$, $k=1,2,\ldots$ be a sequence of random variables. We denote by $\sigma(X_n, X_{n+1},\ldots)$ the sigma algebra generated by $\{X_k\}_{k\geq n}$. The tail sigma algebra of the process $\{X_n\}_{n\geq 1}$ is defined by ${\cal T}= \cap_{n\geq 1} \sigma(X_n, X_{n+1},\ldots)$.

The following result is a generalization of the Kolmogorov zero-one law, valid for a process $\{X_n\}$ where the random variables are not independent, but have decaying correlations.

\begin{thm}[Extended Kolmogorov zero-one law]
\label{01law}
Let $X_n$ be a sequen\-ce of random variables with decaying correlations, as in \fer{016}. Then we have $P(A)=0$ or $P(A)=1$ for any tail event $A\in{\cal T}$. 
\end{thm}

A proof of this result can be obtained by extending proofs of the Kolmogorov zero-one law for independent variables, see \cite{A}. 
Under Condition A, we write the rank-one Riesz projection of $T$ associated to $z=1$ as
\begin{equation}
\label{mm-1}
\Pi = |\psi_\s\rangle\langle\psi_\s^*|,
\end{equation}
where $T\psi_\s=\psi_\s$, $T^*\psi_\s^*=\psi_\s^*$ and $\scalprod{\psi_\s^*}{\psi_\s}=1$, $\|\psi_\s\|=1$. ($T^*$ is the adjoint of $T$.) 
\begin{trace}
Assume that Condition A holds and let $S\subset {\rm spec}(M)$. Then
$$
P(X_n\in S \mbox{\ eventually}) = \scalprod{\psi_\s^*}{\Pi_S\psi_\s}  \in \{0,1\}.
$$ 
\end{trace}

{\it Remark.\ } Both $\psi_\s^*$ and $\Pi_\s$ depend on $\lambda$ ($\psi_\s$ does not). If condition A holds for $\lambda\in I\backslash\{0\}$ for some neighbourhood $I\subset\mathbb R$ of zero, then for $\lambda$ sufficiently small we have $P(X_n\in S \mbox{\ eventually}) =\lim_{\lambda\rightarrow 0} \scalprod{\psi_\s^*(\lambda)}{\Pi_S(\lambda)\psi_\s}$. 
This follows from the facts that the mapping $\lambda\mapsto \scalprod{\psi_\s^*}{\Pi_S\psi_\s}$ is continuous in a deleted neighbourhood of $\lambda=0$ and that the image is discrete. We point out that the map $\lambda\mapsto \scalprod{\psi_\s^*}{\Pi_S\psi_\s}$ is actually holomorphic in a punctured neighbourhood of the origin, and stays bounded there (the only possible image points being $0$ or $1$, even for complex $\lambda$, by the identity principle). Hence zero is a removable singularity of this map. As objects on their own, $\psi_\s^*$ and $\Pi_\s$ are not holomorphic at the origin in general (eigenvalue splitting), but their combination as in the inner product is.

\medskip
Theorem \ref{propx1} gives a criterion for $P(X_n\in S \mbox{\ eventually})=0$. The next result is a characterization of when this probability is one. 
\begin{lem}
\label{lemma100} 
Assume that Condition A holds. If $\scalprod{\psi_\s}{\Pi_S\psi_\s}\neq 0$ for some small enough $\lambda$ then $P(X_n\in S \mbox{\ eventually})=1$ for sufficiently small values of $\lambda$.
\end{lem}
{\it Proof of Lemma \ref{lemma100}.\ } The inequality $P(X_n\in S \mbox{\ eventually})\geq P(X_n\in S\ \forall n\geq 1)$ is the same as $\scalprod{\psi_\s}{\Pi\,  \Pi_S\psi_\s}\geq \scalprod{\psi_\s}{\Pi_S\psi_\s}$ (see also Appendix A). The result now follows from the fact that the left side can only take the values zero or one, independently of $\lambda$ for $\lambda$ sufficiently small (guaranteeing that $\scalprod{\psi_\s}{\Pi\,  \Pi_S\psi_\s}$ is continuous). \hfill $\blacksquare$

\begin{prop}
\label{propx2} 
{}Let $A_j\in\sigma(X_j)$, $j\geq 1$. We have 
$$
\sup_{n\geq 1}\left|P(A_n,\ldots, A_{n+k})-P(A_n)\cdots P(A_{n+k})\right| \leq C_k \|V\|
$$ 
for any $k\geq 1$, and  for some constant $C_k$.
\end{prop}

{\it Proof of Proposition \ref{propx2}.\ }
It suffices to show that 
$$
P(X_n\in S_n,\ldots,X_{n+k}\in S_{n+k})-P(X_n\in S_n)\cdots P(X_{n+k}\in S_{n+k}) =O(\|V\|),
$$
uniformly in $n$, and where $S_j\subseteq{\rm spec}(M)$. Since $T_S = PB^*B\e^{\i\tau K}E_SP = \e^{\i\tau L_\s} \omega_{\rm in}(E_S)+O(\|V\|)$, the joint probability on the left side is
$$
\scalprod{\psi_\s}{T^{n-1} T_{S_n}\cdots T_{S_{n+k}}\psi_\s} = \omega_{\rm in}(E_{S_n})\cdots\omega_{\rm in}(E_{S_{n+k}})+O(\|V\|).
$$ 
Similarly, $P(X_j\in S_j)=\omega_{\rm in}(E_{S_j})+O(\|V\|)$, and so the result follows.\hfill $\blacksquare$

\medskip
 Let $\omega_n$ be the state of $\s$ at time step $n$ (obtained by reducing the state of the entire system to $\s$). 

\begin{lem}[Evolution of averaged system state]
\label{lemma102}
The system state at time step $n$, $\omega_n$, is a random variable (determined by the random measurement history). Its expectation, ${\mathbb E}[\omega_n]$, equals the state obtained by evolving the initial condition according to the dynamics {\em without} measurement.
\end{lem}

{\em Proof of Lemma \ref{lemma102}.\ } For a given measurement path $X_1=m_1,\ldots,X_n=m_n$, the system state is
$$
\omega_n(A) = \frac{\scalprod{\psi_\s}{T_1\cdots T_n A\psi_\s}}{\scalprod{\psi_\s}{T_1\cdots T_n\psi_\s}},
$$
where $A$ is any system observable and $T_j= T_{\{m_j\}}$. Since $P(X_1=m_1,\ldots, X_n=m_n)=\scalprod{\psi_\s}{T_1\cdots T_n\psi_\s}$, this yields
\begin{eqnarray*}
{\mathbb E}[\omega_n(A)] &=& \sum_{m_1,\ldots,m_n} P(X_1=m_1,\ldots, X_n=m_n)\frac{\scalprod{\psi_\s}{T_1\cdots T_n A\psi_\s}}{\scalprod{\psi_\s}{T_1\cdots T_n\psi_\s}} \\
&=&  \sum_{m_1,\ldots,m_n}\scalprod{\psi_\s}{T_1\cdots T_n A\psi_\s}\\
&=& \scalprod{\psi_\s}{T^n A\psi_\s}.
\end{eqnarray*}
In the last step, we have used that 
$$
\sum_m T_{\{m\}}=\sum_m PB^*B\e^{\i\tau K}E_{\{m\}}P=PB^*B\e^{\i\tau K}P=T.
$$
The right hand side is the single-step dynamics operator of the system without probe measurements. \hfill $\blacksquare$

\medskip

Lemma \ref{lemma102} shows in particular that the expectation of the system state converges to the repeated interaction state, 
$$
\lim_{n\rightarrow\infty}{\mathbb E}[\omega_n(A)]=\scalprod{\psi_\s^*}{(A\otimes\bbbone_\s)\psi_\s},
$$
see also \fer{mm-1}. This, of course, does not mean that $\omega_n$ itself converges, in general. However, if the measurement process converges, then we do have the following result.

\begin{lem}[Asymptotic state of $\s$]
\label{lemma101} 
Suppose that the measurement outcomes 
\begin{equation}
\label{outcomes}
X_1\in S_1,\ldots, X_{n-1}\in S_{n-1},\quad X_k\in S, \ k\geq n
\end{equation}
are observed for some $n\geq 1$, and that 1 is a simple eigenvalue of $T_S$ with  Riesz projection $\Pi_S=|\psi\rangle\langle\psi^*|$. We have for any observable $A$ of $\s$ 
$$
\lim_{n\rightarrow\infty}\omega_n(A) = \omega_\infty(A)=\frac{\scalprod{\psi^*}{A\psi_\s}}{\scalprod{\psi^*}{\psi_\s}}.
$$
\end{lem}

\bigskip
This is an inverse scattering result: knowing that the scattered particles are measured to lie in $S$ we can deduce the state of the scattering object $\s$. The final state does not depend on the initial outcomes $X_n$ for $n<k$, any $k$. However we show in the proof of Lemma \ref{lemma101} that if the eigenvalue 1 of $T_S$ is not simple, then the system converges to a final state which depends on the whole measurement path $X_1, X_2,\ldots$

\medskip

{\it Proof of Lemma \ref{lemma101}.\ } The asymptotic state of the system is
$$
\omega_\infty(A) =\lim_{l\rightarrow\infty}\frac{\scalprod{\psi_\s}{T_1\cdots T_{n-1}T_S^l A\psi_\s}}{\scalprod{\psi_\s}{T_1\cdots T_{n-1}T_S^l\psi_\s}},
$$
see \fer{mm15}. Now $\scalprod{\psi_\s}{T_1\cdots T_{n-1}T_S^l\psi_\s}$ converges to the nonzero probability of observing \fer{outcomes}. Therefore, by (the proof of) Lemma \ref{lemma10}
$$
\lim_{l\rightarrow\infty}\scalprod{\psi_\s}{T_1\cdots T_{n-1}T_S^l\psi_\s}= \scalprod{\psi_\s}{T_1\cdots T_{n-1}\widetilde\Pi_S\psi_\s}=\scalprod{\psi_\s}{T_1\cdots T_{n-1}\Pi_S\psi_\s}.
$$
By simplicity of the eigenvalue $1$ of $T_S$,
$$
\omega_\infty(A) =\frac{\scalprod{\psi_\s}{T_1\cdots T_{n-1}\psi}\scalprod{\psi^*}{A\psi_\s}}{\scalprod{\psi_\s}{T_1\cdots T_{n-1}\psi}\scalprod{\psi^*}{\psi_\s}}.
$$
If 1 is not a simple eigenvalue of $T_S$, so that $\Pi_S=\sum_{j=1}^r |\psi_j\rangle\langle\psi^*_j|$, the final state is
$$
\omega_\infty(A) =\frac{\sum_{j=1}^r \scalprod{\psi_\s}{T_1\cdots T_{n-1}\psi_j}\scalprod{\psi^*_j}{A\psi_\s}}{\sum_{i=1}^r \scalprod{\psi_\s}{T_1\cdots T_{n-1}\psi_i}\scalprod{\psi^*_i}{\psi_\s}}.
$$
The final state then depends on the whole measurement history.\hfill $\blacksquare$

\section{The truncated Jaynes-Cummings model}
\label{example}

We consider a simple system where both the scatterer and the probes have only two degrees of freedom participating in the scattering process. The pure state space of $\s$ and $\p$ is ${\mathbb C}^2$, and the Hamiltonians are given by the Pauli $\sigma_z$ operator,
\begin{eqnarray}
H_\s=H_\p &=&
 \left[
\begin{array}{cc}
1 & 0\\
0 & -1
\end{array}
\right].
\label{009}
\end{eqnarray}
The interaction between $\s$ and $\p$ is determined by the operator 
\begin{equation}
\lambda V= \lambda\left( a^*_\s\otimes a_\p + a_\s\otimes a_\p^*\right),
\label{0010}
\end{equation}
with coupling constant $\lambda\in {\mathbb R}$, and where  
\begin{equation}
a= \left[
\begin{array}{cc}
0 & 0\\
1 & 0
\end{array}
\right],\qquad
a^*= \left[
\begin{array}{cc}
0 & 1\\
0 & 0
\end{array}
\right]
\label{0011}
\end{equation}
are the annihilation and creation operators. In the usual Jaynes-Cummings model (used e.g. in quantum optics), the system $\s$ has {\em infinitely} many levels (harmonic oscillator), see e.g. \cite{NC} and also \cite{BrPi}. Our model is a truncation, but it still describes energy exchange between $\s$ and $\p$. In what follows, we can treat all values of $\lambda$, not necessarily small ones only. This is so since the model is essentially exactly solvable. The total Hamiltonian $H=H_\s+H_\p+\lambda V$ describes exchange of energy between $\s$ and $\p$, while the total number of excitations, $N=a_\s^*a_\s+a_\p^*a_\p$, is conserved (commutes with $H$). This allows for a treatment of the system separately in the invariant sectors $N=0,1,2$.

{}For an arbitrary probe observable $X\in{\cal B}({\mathbb C}^2)$ we write $X_{ij}=\scalprod{\varphi_i}{X\varphi_j}$, where $\varphi_1$, $\varphi_2$ are the orthonormal eigenvectors of $H_\p$ (with $H_\p\varphi_1=\varphi_1$). Incoming states are determined by $p\in[0,1]$ via 
\begin{equation}
\omega_{\rm in}(X)=p X_{11} +(1-p) X_{22},
\label{nm2}
\end{equation}
where $X\in{\cal B}({\mathbb C}^2)$ is an arbitrary probe observable.

We will use the notation and definitions of Section \ref{qdssetup} in what follows. In particular, the single step operator $T_S$ is defined in \fer{mm14}. For the following explicit formula, we take the reference state $\Psi_\s$ to be the trace state.

\begin{thm}[Explicit reduced dynamics operator]
\label{thmexplicit}
Set $\varphi_{ij}=\varphi_i\otimes\varphi_j$ and let $X$ be any operator of $\p$. In the basis $\{\varphi_{11}, \varphi_{12}, \varphi_{21}, \varphi_{22}\}$ we have 
\begin{eqnarray}
\lefteqn{PB^*B \e^{\i \tau K} XP = \omega_{\rm in}(X) \,\e^{\i\tau L_\s}+}\label{nm1}\\
&&\qquad \small\nonumber\\
&&\left[
\small
\begin{array}{cc}
(1-p) X_{22} a & (1-p)X_{21}b \nonumber\\
-pX_{12}\e^{2\i\tau} \i \sin(\lambda\tau) & \e^{2\i\tau}(\cos(\lambda\tau)-1)\omega_{\rm in}(X) \nonumber\\
pX_{21}\e^{-2\i\tau}\i\sin(\lambda\tau) & 0 \nonumber\\
-pX_{22}a & -pX_{21} b
\end{array}
\right. \nonumber\\
&&\qquad \small\nonumber\\
&&\qquad\small
\left.
\begin{array}{cc}
-(1-p)X_{12} b & -(1-p)X_{11} a \nonumber\\
0 & (1-p)X_{12}\e^{2\i\tau}\i\sin(\lambda\tau)\nonumber\\
\e^{-2\i\tau}(\cos(\lambda\tau)-1)\omega_{\rm in}(X)& -(1-p)X_{21}\e^{-2\i\tau}\i\sin(\lambda\tau)\nonumber\\
pX_{12} b & pX_{11} a\nonumber
\end{array}
\right],
\end{eqnarray}
where $a=-\sin^2(\lambda\tau)$, $b=-\i\sin(\lambda\tau)\cos(\lambda\tau)$.
\end{thm}
We point out that the vector $[p, 0, 0, 1-p]^t$ is an eigenvector of the adjoint of \fer{nm1} with eigenvalue $\omega_{\rm in}(X)$.

\medskip

{\em Proof.\ } The proof is obtained by an explicit calculation. Since $H_\s$ and $H_\p$ commute with $I=V-V'$, where $V'= J\Delta^{1/2}V\Delta^{-1/2}J$ (see \fer{mm21}), it suffices to calculate
$$
PB^*B \e^{\i\tau I} XP = \sum_{n=0}^\infty \frac{(\i\tau)^n}{n!}\sum_{k=0}^n {n\choose k} (-1)^{n-k} PB^*B V^k(V')^{n-k} XP.
$$
Here it is understood that all operators are considered in the ``doubled'' (GNS) Hilbert space, e.g., 
$$
V = a^*_\s\otimes\bbbone_\s \otimes a_\p\otimes\bbbone_\p + a_\s\otimes\bbbone_\s \otimes a^*_\p\otimes\bbbone_\p.
$$
Powers of $V$ and $V'$ can be calculated explicitly. For instance, for $k\geq 2$ even, we have $V^k = \widehat n^{k/2}\otimes\bbbone\otimes (1-\widehat n)^{k/2}\otimes\bbbone + (1-\widehat n)^{k/2}\otimes\bbbone\otimes\widehat n^{k/2}\otimes\bbbone$, where $\widehat n=a^*a$. One obtains similar expressions for $k$ odd, and for $(V')^l$. Using these expression in the above series, one gets the result of Theorem \ref{thmexplicit}.\hfill $\blacksquare$

\bigskip

{\bf Resonant and non-resonant system.\ } If $\lambda\tau$ is a multiple of $\pi$ then \fer{nm1} reduces to $PB^*B\e^{\i\tau K}XP= \omega_{\rm in}(X)\, {\rm diag}(1,\pm 1,\pm 1,1)$ with plus and minus signs if the multiple is even and odd, respectively. Then, by using the expression $P(X_1\in S_1,\ldots,X_n\in S_n)$ given in \fer{mm16}, it is readily seen that the random variables $X_j$ are independent, and $P(X_j\in S)=\omega_{\rm in}(E_S)$. When  $\lambda\tau\in \pi{\mathbb Z}$ we call the system {\it resonant} \cite{BrPi}, otherwise we call it {\it non-resonant}. One can understand the resonant regime as follows: consider the dynamics generated on $\s$ and a single probe $\p$ by the Hamiltonian $H=H_\s+ H_\p + \lambda V$. The probability of transition from the initial state $\varphi_2^\s\otimes\varphi_1^\p$, where the $\s$ is in the ground state and $\p$ in the excited state, to the opposite state $\varphi_1^\s\otimes\varphi_2^\p$, at time $t$, is given by $P_t = \left|\scalprod{\varphi_1^\s\otimes\varphi_2^\p}{\e^{\i tH}\varphi_2^\s\otimes\varphi_1^\p}\right|^2= \sin^2(\lambda t)$. For  $\lambda t\in \pi{\mathbb Z}$ this probability vanishes. If the interaction time $\tau$ in the repeated interaction system is a multiple of $\pi/\lambda$, then interaction effects are suppressed. It is not hard to see that in this case, the system does not feel the interaction with the probes in the sense that $\omega_n(A)=\omega_0(\alpha^\s_n(A))$ for all $n\geq 1$, where $\alpha^\s_n(A)$ is the reduced dynamics of $\s$ alone. We focus now on the non-resonant situation.

\medskip

{\bf Asymptotics of the measurement process.\ }  We suppose the incoming state of the probes is given by $\omega_{\rm in}=|\varphi_1\rangle\langle\varphi_1|$, i.e., they are in the pure spin-up state. This corresponds to $p=1$ in \fer{nm2}. Let $M$ be a measurement operator, $S\subset{\rm spec}(M)$ and let $E_S$ be the projection onto the corresponding spectral subspace. The operator \fer{nm1} with $X=E_S$ has spectrum 
\begin{equation}
\label{xxx1}
{\rm spec}(T_S) = (E_S)_{11}\ \big\{ 1, \e^{2\i\tau}\cos(\lambda\tau), \e^{-2\i\tau}\cos(\lambda\tau), \cos^2(\lambda\tau)\big\}.
\end{equation}
Since $E_S$ is a projection, we have $0\leq (E_S)_{11}=\scalprod{\varphi_1}{E_S\varphi_1}\leq 1$. 

$\circ$ The equality $(E_S)_{11}=1$ holds if and only if $E_S\psi_1=\psi_1$, so if and only if either $E_S=\bbbone$ or $E_S=|\varphi_1\rangle\langle\varphi_1|$. The case $E_S=\bbbone$ is discarded since it corresponds to not making a measurement. Hence if $(E_S)_{11}=1$ in presence of a measurement implies $E_S=|\varphi_1\rangle\langle\varphi_1|$. This forces the measurement operator $M$ to be diagonal in the basis $\{\varphi_1,\varphi_2\}$, i.e., $M={\rm diag}(m_1, m_2)$. Conversely, if $M$ is diagonal and $E_S=|\varphi_1\rangle\langle\varphi_1|$, then $(E_S)_{11}=1$. It follows that $T_S$ has an eigenvalue $1$ if and only if $M=m_1|\varphi_1\rangle\langle\varphi_1| +m_2|\varphi_2\rangle\langle\varphi_2|$ and $E_S=|\varphi_1\rangle\langle\varphi_1|$. In this case, the associated Riesz spectral projection is $\Pi = \sqrt{2} |\psi_\s\rangle\langle\varphi_{11}|$ (see \fer{projections}) and we have $P(X_n=m_1{\rm \ eventually})=1$.

$\circ$ If the measurement operator $M$ is not diagonal in the basis $\{\varphi_1,\varphi_2\}$, then $(E_S)_{11}<1$ for any $S$ with $|S|=1$ (and again, if $|S|=2$ then $E_S=\bbbone$ which means we do not make a measurement). Then $1$ is not an eigenvalue of $T_S$, according to \fer{xxx1}, and so $P(X_n\in S{\rm \ eventually})=0$.  

We conclude that the measurement process converges if and only if the incoming state is pure and localized with respect to the measurement operator (i.e., if and only if it is given by an eigenvector of $M$ and we measure the corresponding eigenvalue). In the situation where the measurement outcomes converge, we can determine the asymptotic state of the scatterer $\s$ from Lemma \ref{lemma101}. It is given by $\omega_\infty(A) =\omega_{\rm in}(A)$, $A\in{\cal B}({\mathbb C}^2)$, thus the state of the incoming probe is copied onto the scatterer.  This copying mechanism has been described before as ``homogenization'' in \cite{BG}. (Our analysis is more complete than previous ones, as it describes the entire system of scatterer and probes.)  Note also that the asymptotic mean is given by $\mu_\infty=\omega_{\rm in}(M)$. The frequencies are $f_m=\omega_{\rm in}(E_m)$ (expression \fer{fmrep}).  This suggests that the cavity becomes `transparent' for large times (no effect on incoming probes). 

\medskip

{\bf Large deviations for the mean.\ }The logarithmic moment generating function \cite{DZ} is defined by 
\begin{equation}
\Lambda(\alpha) = \lim_{n\rightarrow\infty}\frac1n \log\E[\e^{n\alpha\, \overline{\!X}_{\! n}}],
\label{001}
\end{equation}
for $\alpha\in\mathbb R$ s.t. the limit exists as an extended real number. Using expression \fer{Lambda} and Theorem \ref{thmexplicit} (with  $p=1$), we find that $\Lambda(\alpha)=\log\omega_{\rm in}(\e^{\alpha M})$, for $\alpha\in\mathbb R$. The Legendre transformation of $\Lambda(\alpha)$,
\begin{equation}
\Lambda^*(x) = \sup_{\alpha\in{\mathbb R}} \ \alpha x-\Lambda(\alpha),
\label{b1}
\end{equation}
$x\in\mathbb R$, is called the {\it rate function}. Its usefulness in the present context is due to the G\"artner-Ellis theorem \cite{DZ}, which asserts that for any closed set $F\subset\mathbb R$ and any open set $G\subset\mathbb R$, we have
\begin{eqnarray}
\limsup_{n\rightarrow\infty} \frac1n \log P\left(\,\overline{\! X}_{\! n} \in F \right)&\leq& -\inf_{x\in F}\Lambda^*(x)\nonumber \\
\liminf_{n\rightarrow\infty} \frac1n \log P\left(\, \overline{\! X}_{\! n}\in G\right)&\geq &-\inf_{x\in G\cap{\cal F}}\Lambda^*(x).
\label{b10}
\end{eqnarray}
Here, $\cal F$ denotes the set of `exposed points of $\Lambda^*$' (see \cite{DZ}). 

\begin{prop}
\label{bprop}
Suppose that ${\rm Var}(M)=\omega_{\rm in}(M^2)-\omega_{\rm in}(M)^2$, the variance of $M$ in the state $\omega_{\rm in}$, does not vanish. Then $\Lambda^*$ is holomorphic at $x=\omega_{\rm in}(M)$, and 
$$
\Lambda^*(x) = \frac{\left( x-\omega_{\rm in}(M)\right)^2}{2{\rm Var}(M)} +O\!\left( (x-\omega_{\rm in}(M))^4\right).
$$
\end{prop}

{\em Proof.\ } Note that $\Lambda$ is twice differentiable, and the second derivative w.r.t. $\alpha$ of the argument of the supremum in \fer{b1} is less than or equal to zero. Therefore, for fixed $x$, the supremum is taken at $\alpha\in\mathbb R$ satisfying 
\begin{equation}
x = \Lambda'(\alpha)=\frac{\omega_{\rm in}(M\e^{\alpha M})}{\omega_{\rm in}(\e^{\alpha M})}.
\label{b2}
\end{equation}
For $\alpha=0$ we have $x=\omega_{\rm in}(M)$. If $\Lambda''(0)= {\rm Var}(M)\neq 0$, then equation \fer{b2} has an implicit solution $\alpha=\alpha(x)$, locally around $x=\omega_{\rm in}(M)$.  Since $\Lambda'(\alpha)$ is holomorphic at $\alpha=0$, the implicit solution  is holomorphic at $x=\omega_{\rm in}(M)$ (see e.g. \!\cite{SZ}, p.163, equation (12.4)). The Taylor expansion of \fer{b2} is 
\begin{equation}
x=\omega_{\rm in}(M) +\alpha{\rm Var}(M) +c\alpha^2 +O(\alpha^3),
\label{b3}
\end{equation}
where $c=\frac12\{ \omega_{\rm in}(M^3)-3\omega_{\rm in}(M^2)\omega_{\rm in}(M)+2\omega_{\rm in}(M)^3\}$. We solve equation \fer{b3} implicitly for $\alpha=\alpha(x)$, which is the point where the supremum in \fer{b1} is taken. The explicit formula for the supremum given in Proposition \ref{bprop} follows.\hfill $\blacksquare$

\bigskip

{\it Example: Measuring the outgoing spin angle.\ }
Since $\omega_{\rm in}$ is the state `spin up', we have $\omega_{\rm in}(M)=M_{11}$ and ${\rm Var}(M)= |M_{12}|^2$. Imagine an experiment where we measure the angle of the spins as they exit the scattering process. Let $\theta\in[0,\pi/2)$ and $\phi\in[0,2\pi)$ be the angles measuring the altitude ($\theta=0$ is spin up) and azimuth ($\phi=0$ is the plane orthogonal to the axis of the cavity). The measurement operator ``spin in direction $(\theta,\phi)$'' is given by
$$
M=\left[
\begin{array}{cc}
\cos\theta & \e^{-\i\phi} \sin\theta\\
\e^{\i\phi} \sin\theta & -\cos\theta
\end{array}
\right],
$$
see e.g. \cite{CT}, Chapitre IV, (A-19). The eigenvectors of $M$ associated to the eigenvalues $\pm 1$ of $M$ are 
\begin{eqnarray*}
\chi_+ &=& \e^{-\i\phi/2}\cos(\theta/2) \varphi_1 + \e^{\i\phi/2}\sin(\theta/2) \varphi_2 \\
\chi_- &=& -\e^{-\i\phi/2}\sin(\theta/2) \varphi_1 + \e^{\i\phi/2}\cos(\theta/2) \varphi_2.
\end{eqnarray*}
The eigenprojection $E_+$ measures the spin in the positive direction $(\theta,\phi)$. By using Lemma \ref{lem3} is easy to see that
$$
P\left( \mbox{$X_n$ is in direction $(\theta,\phi)$ eventually}\right) = \left\{
\begin{array}{ll}
1 & \mbox{if $\theta =0$}\\
0 & \mbox{if $\theta\neq 0$.}
\end{array}
\right.
$$
This is another manifestation of the asymptotic transparency of the cavity.

We obtain from theorem \ref{prop01'} (with $\mu_\infty=\cos\theta$) that for any $\epsilon>0$,  
$$
\lim_{n\rightarrow\infty} P(|\,\overline{\! X}_{\! n}-\cos\theta|\geq \epsilon)=0.
$$
The speed of convergence can be estimated using \fer{b10} and Proposition \ref{bprop}. It is easy to see that the logarithmic generating function and the rate function associated to the shifted random variable $\overline{\!X}_{\! n}-\cos\theta$ are given by $\Lambda_{\rm shift}(\alpha) = \Lambda(\alpha)-\alpha\cos\theta$ and  $\Lambda_{\rm shift}^*(x)=\Lambda^*(x+\cos\theta)$, respectively. Next, we note that all points in the vicinity of zero belong to the set ${\cal F}_{\rm shift}$, the set of exposed points of $\Lambda^*_{\rm shift}$. Indeed, if $x=\Lambda_{\rm shift}'(\alpha)$ for some $\alpha\in{\mathbb R}$, then $x\in{\cal F}_{\rm shift}$ (\cite{DZ}, Lemma 2.3.9). But $x=0=\Lambda_{\rm shift}'(0)$, and $\Lambda'_{\rm shift}$ is invertible around zero (as $\Lambda''_{\rm shift}(0)\neq 0$). This shows that ${\cal F}_{\rm shift}$ contains a neighbourhood of the origin.

Take $0<\epsilon<\epsilon'<\!\!<1$, set $G=(-\epsilon',-\epsilon)\cup (\epsilon,\epsilon')$, and let $F$ be the closure of $G$. Then $\inf_{x\in F}\Lambda_{\rm shift}^*(x) = \inf_{x\in G\cap{\cal F}_{\rm shift}}\Lambda_{\rm shift}^*(x) = \frac{\epsilon^2}{2{\rm Var}(M)} +O((\epsilon')^4)$. (We use Proposition \ref{bprop}.) Combining this with the two bounds \fer{b10} (for the shifted random variable), we obtain
$$
P\big(\epsilon \leq |\,\overline{\! X}_{\! n}-\cos\theta|\leq \epsilon')\sim \exp\left[-n\Big\{\frac{\epsilon^2}{2\sin^2\theta} + O((\epsilon')^4)\Big\}\right],\quad n\rightarrow\infty,
$$
which is a large deviation statement for the average $\,\overline{\! X}_{\! n}$.

\section{Proofs}

\subsection{Proof of Theorems \ref{corrdeclemma'} and \ref{corrdeclemma}} 
\label{proof1+5}

Theorem \ref{corrdeclemma} is a stronger version of Theorem \ref{corrdeclemma'}, so it suffices to prove the former.

Let $A \in \sigma(X_k,\ldots, X_l)$ and $B\in \sigma(X_m,\ldots, X_n)$. The range $\Sigma$ of the $X_n$ is finite, so $\sigma(X_k,\ldots, X_l)$ consists of the collection of all sets of the form $\{\omega : (X_k(\omega),\ldots,X_l(\omega))\in H\}$, where $H\subseteq \Sigma^{l-k+1}$ (\cite{B}, Thm. 5.1). Therefore we have
\begin{eqnarray*}
 A &=& \bigcup_{j=1}^J X_k^{-1}(\{s_k^{(j)}\}) \cap \cdots \cap X_l^{-1}(\{s_l^{(j)}\}) =: \bigcup_{j=1}^J A_j \\
B &=& \bigcup_{i=1}^I X_m^{-1}(\{s_m^{(i)}\})\cap \cdots \cap X_n^{-1}(\{s_n^{(i)}\}) =: \bigcup_{i=1}^I B_i,
\end{eqnarray*}
where $s_r^{(j)}, s_r^{(i)}\in \Sigma$, and $A_i \cap A_j = \emptyset$, $B_i \cap B_j =\emptyset$ for $i\neq j$. Thus,
\begin{equation*}
P(A)=\sum_j P(A_j),\  P(B)=\sum_i P(B_i), \  P(A\cap B) = \sum_{i,j}P(A_j \cap B_i).
\end{equation*}
Setting $T_m^{(n)}:= T_{S=\{s_m^{(n)}\}}$, we have
\begin{equation}
P(A\cap B) = \sum_{i,j} \langle \Psi_S, T^{k-1} T_k^{(j)}\cdots T_l^{(j)}T^{m-l-1} T_m^{(i)}\cdots T_n^{(i)} \Psi_S \rangle.
\label{012.1}
\end{equation}
We now approximate $T^{m-l-1}$ by its value for large $m-l$. To do so, let $P_\s$ denote the Riesz spectral rank-one projection onto the eigenvalue one of $T$, and let $\pbar_\s =\bbbone_\s-P_\s$. We have $P_\s=|\psi_\s\rangle\langle\psi_\s^*|$, where $\psi_\s^*\in\ch_\s$ satisfies $T^*\psi^*_\s=\psi^*_\s$ and $\scalprod{\psi^*_\s}{\psi_\s}=1$. The operator $T$ has the spectral representation \cite{K}, I\textsection 5
\begin{equation}
T = |\psi_\s\rangle\langle\psi_\s^*| + \sum_{r=1}^d \{ z_r P_r +D_r\},
\label{001'}
\end{equation}
where $P_r$ is the Riesz projection associated to the eigenvalue $z_j$ and $D_r$ is the associated eigen-nilpotent. We have $D_r^{\nu_r}=0$ ($\nu_r$ is the index of $z_r$). Consequently,
\begin{equation}
T^k = |\psi_\s\rangle\langle\psi_\s^*| + \sum_{r=1}^d \sum_{q=0}^{\nu_r-1} {k\choose q} z_r^{k-q} P_r D_r^q=:|\psi_\s\rangle\langle\psi_\s^*| + R_k.
\label{011}
\end{equation}
Note that it suffices to consider the nonzero $z_r$ in \fer{011}. 
Using the bound $\sum_{q=0}^{\nu-1}{k\choose q}\leq \nu k^{\nu-1}\leq k^\nu$, and the fact that for any $\epsilon>0$ there is a constant $C_\epsilon$ s.t. $k^\nu\leq C_\epsilon\e^{\epsilon k}$ for all $k\geq 1$, we obtain 
\begin{equation}
\|R_k\| \leq C_\epsilon\e^{\epsilon k}\max_{1\leq r\leq d} |z_r|^k\leq C_\epsilon\e^{\epsilon k}(1-\gamma)^k, 
\label{002}
\end{equation}
where we invoke Assumption A, $|z_r|\leq 1-\gamma$.
We now replace $T^{m-l-1}$ in \fer{012.1} using \fer{011},
\begin{eqnarray}
P(A\cap B)&=& P(A)\sum_{i} \langle \psi_S^*, T_m^{(i)}\cdots T_n^{(i)} \psi_S \rangle \label{013-1}\\
&& + \sum_{i,j} \langle \psi_S,T^{k-1} T_k^{(j)}\cdots T_l^{(j)}R_{m-l-1} T_m^{(i)}\cdots T_n^{(i)} \psi_S \rangle 
\label{013}
\end{eqnarray}
The sum in \fer{013-1} equals 
\begin{eqnarray}
\sum_i \langle \psi_S^*, T_m^{(i)}\cdots T_n^{(i)} \psi_S \rangle &=& 
\sum_i \langle \psi_\s, \big( |\psi_\s\rangle\langle\psi_\s^*|\big) T_m^{(i)}\cdots T_n^{(i)} \psi_S \rangle\nonumber\\
&=& \sum_i \langle \psi_\s, \big(T^{m-1}-R_{m-1}\big) T_m^{(i)}\cdots T_n^{(i)} \psi_\s \rangle\nonumber\\
&=& P(B) - \sum_i \langle \psi_\s, R_{m-1} T_m^{(i)}\cdots T_n^{(i)} \psi_\s \rangle.
\label{--1}
\end{eqnarray}
\begin{lem}
\label{lemma05}
There is a constant $C$ s.t. $\|\sum_i T_m^{(i)}\cdots T_n^{(i)}\|\leq C$, independently of the range of values of $i,m,n$ and of the $s_l^{(i)}$ defining the $T_l^{(i)}$.
\end{lem}

We give a proof below. Lemma \ref{lemma05} together with \fer{013-1}, \fer{013} and \fer{--1} gives
\begin{eqnarray}
\lefteqn{
\left| P(A\cap B)-P(A)P(B)\right|\leq C P(A)\|R_{m-1}\|}\nonumber\\
&&+\left| \sum_{i,j} \langle \psi_\s,T^{k-1} T_k^{(j)}\cdots T_l^{(j)}R_{m-l-1} T_m^{(i)}\cdots T_n^{(i)} \psi_\s \rangle \right|.
\label{--2}
\end{eqnarray}
The remainder $R_{m-l-1}$ is given by the sum in \fer{011} (with $k=m-l-1$). We expand
\begin{equation}
P_rD_r^q = \sum_{s,s'} \omega(s,s')|\chi_s\rangle\langle\chi_{s'}|
\label{003}
\end{equation}
in an orthonormal basis $\{\chi_s\}$, where $\omega(s,s')\in\mathbb C$ are matrix elements (also depending on $r,q$). The modulus of the sum in \fer{--2} is bounded above by
\begin{eqnarray}
\lefteqn{
C \sum_{r=1}^d\sum_{q=0}^{\nu_r-1} {m-l-1\choose q} |z_r|^{m-l-1-q} \sum_{s,s'}|\omega(s,s')|}\nonumber\\
&&\times \sum_{j}\left| \langle\psi_\s, T^{k-1} T^{(j)}_k\cdots T_l^{(j)} \chi_s\rangle\right|.
\label{--4}
\end{eqnarray}
\begin{lem}
\label{lemma06}
Let $A_s\in{\frak M}_\s$ be the unique operator s.t. $\chi_s=A_s\psi_\s$. We have 
$$
\sum_{j}\left| \langle\psi_\s, T^{k-1} T^{(j)}_k\cdots T_l^{(j)} \chi_s\rangle\right|\leq \|A_s\| P(A).
$$
\end{lem}
We give a proof of Lemma \ref{lemma06} below. Using the result of the Lemma in \fer{--4} we obtain, for any $\epsilon>0$, the upper bound  $P(A) C_\epsilon \e^{\epsilon (m-l)} (1-\gamma)^{m-l}$ for the sum in \fer{--2}. Thus for all $\epsilon >0$ there is a $C_\epsilon$ s.t. 
$$
\left| P(A\cap B)-P(A)P(B)\right|\leq C_\epsilon P(A)\big\{ (1-\gamma)^m \e^{\epsilon m} + (1-\gamma)^{m-l}\e^{\epsilon(m-l)}\big\}.
$$
Since $m\mapsto (1-\gamma)^m\e^{\epsilon m}$ is decreasing (for $\epsilon$ small enough), we get \fer{010}. This concludes the proof of Theorem \ref{corrdeclemma} modulo proofs of Lemmata \ref{lemma05} and \ref{lemma06}.

\medskip
{\it Proof of Lemma \ref{lemma05}.\ }
Any vector $\psi\in{\cal H}_\s$ is of the form $\psi = \tilde A\Psi_\s$, for some $\tilde A\in\cm_\s$. We have 
\begin{eqnarray*}
\lefteqn{\left\| \sum_i T_m^{(i)}\cdots T_n^{(i)} \tilde{A} \Psi_\s \right\| =\sup_{\| \tilde{B}\Psi_\s\|=1} \left| \langle \tilde{B}\Psi_\s, \sum_i T_m^{(i)}\cdots T_n^{(i)} \tilde{A} \Psi_\s \rangle \right| }\\
& =& \sup_{\| \tilde{B}\Psi_\s\|=1} | \langle \tilde{B} \Psi_\s \otimes B \Psi_{{\cal P}_m} \otimes\cdots \otimes B \Psi_{{\cal P}_n}, \e^{i \tau L_n}\cdots \e^{i \tau L_m} Q  \tilde{A} \times \\
& & \times \e^{-i \tau L_m}\cdots \e^{-i \tau L_n} \Psi_\s \otimes B \Psi_{{\cal P}_m}\otimes\cdots \otimes B \Psi_{{\cal P}_n} \rangle |,
\end{eqnarray*}
where $Q:= \sum_i E_m(\{s_m^{(i)}\})\cdots E_n(\{s_n^{(i)}\})$ is a selfadjoint projection, $\|Q\|=1$. Thus 
$$
\| \sum_i T_m^{(i)}\ldots T_n^{(i)} \tilde{A} \Psi_s \| \leq \| \tilde{A} \|.
$$
The result now follows from the uniform boundedness principle. This proves Lemma \ref{lemma05}.

\medskip
{\it Proof of Lemma \ref{lemma06}.\ }
By using the definition $T^{(j)}_k= P_k B^*B \e^{\i\tau K}E_k^{(j)}P_k$, where $E_k^{(j)}=E_k(\{s_k^{(j)}\})$, see \fer{mm14}, we see that 
$$
 \langle\psi_\s, T^{k-1} T^{(j)}_k\cdots T_l^{(j)} A_s\psi_\s\rangle = \langle \Psi^{(j)}, A_s\Psi^{(j)}\rangle,
$$
where $\Psi^{(j)}=E^{(j)}_k\cdots E_l^{(j)} \e^{\i\tau L_l}\cdots\e^{-\i\tau L_1}\psi_\s\otimes B\psi_\p\cdots\otimes B\psi_\p$ ($l$ probes). To arrive at the form of $\Psi^{(j)}$, we replace the action of $\e^{\i\tau K}$ in the operators $T_k^{(j)}$ (see just above) by the action of the Liouville operators $\e^{\i\tau L_k} \cdot \e^{-\i\tau L_k}$, see \fer{f11}, \fer{mm13}. Since a positive linear functional on $\cm_\s$ is bounded, with norm equal to its value for the observable $\bbbone$ (\cite{BR} Prop. 2.3.11), we have 
$$
\left|\langle\psi_\s, T^{k-1} T^{(j)}_k\cdots T_l^{(j)} A_s\psi_\s\rangle\right|\leq \|A_s\| \langle\psi_\s, T^{k-1} T^{(j)}_k\cdots T_l^{(j)} \bbbone\psi_\s\rangle.
$$
Note that the scalar product on the r.h.s. is non-negative, as it is a probability. It now follows that
$$
\sum_{j}\left| \langle\psi_\s, T^{k-1} T^{(j)}_k\cdots T_l^{(j)} \chi_s\rangle\right|\leq \|A_s\|\sum_j \langle\psi_\s, T^{k-1} T^{(j)}_k\cdots T_l^{(j)} \psi_\s\rangle,
$$
the latter sum being $P(A)$. \hfill $\blacksquare$

\subsection{Proof of Theorem \ref{propx1}} 

We have $P(X_n\in S \mbox{\ eventually}) = \scalprod{\psi_\s}{\Pi\Pi_S\psi_\s}$, where $\Pi$ and $\Pi_S$ are the Riesz projections of $T$ and $T_S$ respectively, associated to the point $1$, see Lemma \ref{lem3}. We know that $T_S=\omega_{\rm in}(E_S)+O(\|V\|)$. Since $E_S$ is an orthogonal projection and $\omega_{\rm in}$ is a state, $\omega_{\rm in}(E_S)\neq 1$ means $\omega_{\rm in}(E_S)<1$. Thus, for small enough $V$, $T_S$ does not have an eigenvalue at the point $1$, and consequently $\Pi_S=0$.\hfill $\blacksquare$

\subsection{Proof of Theorem \ref{theorem2}}
\label{proofthm2}

The proof is in two parts. First we show that the expectation of $f_m$ has the indicated limit. Then we upgrade this convergence to almost everywhere convergence, using the decay of correlations, Theorem \ref{010'}.

{\it Step 1.} We show convergence of the expectation of the frequency, 
\begin{equation}
F_m:={\mathbb E}[f_m] = \lim_{n\rightarrow\infty} \frac 1n \mathbb{E}\left[\mbox{number of $k\in\{1,\ldots,n\}$ s.t. $X_k=m$}\right].
\label{frequency'}
\end{equation}
Let $m\in{\rm spec}(M)$ be fixed and define $\chi$ by $\chi(x)=1$ if $x=m$ and $\chi(x)=0$ otherwise. Then the expectation in \fer{frequency'} equals
\begin{eqnarray*}
\lefteqn{
\sum_{m_1,\ldots,m_n} \left[\sum_{j=1}^n\chi(m_j)\right] P(X_1=m_1,\ldots,X_n=m_n)}\\
&=& \sum_{j=1}^n\sum_{m_k, k\neq j} P(X_1=m_1,\ldots, X_j=m,\ldots,X_n=m_n)\\
&=&\sum_{j=1}^n \scalprod{\psi_\s}{T^{j-1} T_m\psi_\s}.
\end{eqnarray*}
We have $\frac 1n\sum_{j=0}^{n-1}T^j\rightarrow \Pi$ as $n\rightarrow\infty$ (the ergodic projection which equals the Riesz projection of $T$ associated to one, see Appendix \ref{app}) and, due to Condition A (see \fer{mm35}), $\Pi=|\psi_\s\rangle\langle\psi_\s^*|$, where $\psi_\s^*$ is the unique vector in ${\cal H}_\s$ satisfying $T^*\psi^*_\s=\psi^*_\s$ and $\scalprod{\psi^*_\s}{\psi_\s}=1$. Note that the asymptotic state $\omega_+$ of the dynamics without scattering is $\omega_+(\cdot)=\scalprod{\psi_\s^*}{\cdot\psi_\s}$. We have 
\begin{equation}
F_m = \scalprod{\psi_\s}{\Pi T_m\psi_\s} =\scalprod{\psi^*_\s}{T_m\psi_\s} . 
\label{fmrep}
\end{equation}
Recall that $T_m=PB^*B\e^{\i\tau K}E_mP$, so that 
\begin{eqnarray}
F_m &=& \scalprod{\psi^*_\s\otimes\psi_\p}{(B^*B\e^{\i\tau K}E_m)\psi_\s\otimes\psi_\p}\nonumber\\
&=& \scalprod{\psi^*_\s\otimes\psi_\p}{(B^*B\e^{\i\tau L}E_m\e^{-\i \tau L})\psi_\s\otimes\psi_\p}\nonumber\\
&=& \omega_+\otimes\omega_{\rm in}\big(\e^{\i\tau H} E_m\e^{-\i\tau H}\big).
\label{late1}
\end{eqnarray}
where $L=L_\s+L_\p +\lambda V$, see \fer{f11}.
This finishes the first step of the proof, showing that 
\begin{equation}
\lim_{n\rightarrow\infty} \frac 1n {\mathbb E}\left[\mbox{number of $k\in\{1,\ldots,n\}$ s.t. $X_k=m$}\right] = \omega_+\otimes\omega_{\rm in}\big(\e^{\i\tau H}E_m\e^{-\i\tau H}\big).
\label{lim1}
\end{equation}

{\it Step 2.} We upgrade \fer{lim1} to almost sure convergence, by using a classical fourth moment method. Introduce the random variable
$$
Z_n = \{\mbox{number of $k\in\{1,\ldots,n\}$ s.t. $X_k=m$}\}.
$$
We are going to show below that, for any $\epsilon >0$,
\begin{equation}
\sum_{n\geq 1} P(|Z_n/n -{\mathbb E}[Z_n]/n|^2\geq \epsilon)<\infty.
\label{u1}
\end{equation}
Then by the (first) Borel-Cantelli lemma, $P(|Z_n/n -{\mathbb E}[Z_n]/n|^2\geq \epsilon \mbox{ i.o.})=0$, i.e., there is a set $\Sigma$ of measure one, s.t. for all $\omega\in \Sigma$, there exists a $k$ with
\begin{equation}
|Z_n/n -{\mathbb E}[Z_n]/n|^2 < \epsilon,\qquad \forall n\geq k.
\label{u2}
\end{equation}
{}From \fer{lim1} we know that ${\mathbb E}[Z_n]/n$ converges to $\nu:=\omega_+\otimes\omega_{\rm in}(\e^{\i\tau H}E_m\e^{-\i\tau H})$, so \fer{u2} implies that $Z_n/n$ converges to $\nu$ almost everywhere. It remains to prove the summability \fer{u1}. By Chebyshev's inequality,
\begin{equation}
 P(|Z_n/n -{\mathbb E}[Z_n]/n|^2\geq \epsilon) \leq \frac{1}{\epsilon^2 n^4}{\mathbb E}[|Z_n-{\mathbb E}[Z_n]|^4].
\label{u3}
\end{equation}
We get an upper bound on the r.h.s. Set 
$$
Z_n-{\mathbb E}[Z_n] = \sum_{i=1}^n \{ \chi(X_i=m) - P(X_i=m)\}=:\sum_{i=1}^n Y_i,
$$
so that ${\mathbb E}[Y_i]=0$. Here, $\chi(X_i=m)=1$ if $X_i=m$ and $\chi(X_i=m)=0$ otherwise. We have 
\begin{equation}
{\mathbb E}[| Z_n-{\mathbb E}[Z_n]|^4] \leq \sum_{i,j,k,l=1}^n {\mathbb E}[|Y_i| |Y_j| |Y_k| |Y_l|].
\label{u4}
\end{equation}
The idea is to control the sum by using that if the indices $i,j,k,l$ are far apart from each other, then the expectation is small due to the decay of correlations. Thus only a few terms in the sum contribute to its value. Let $\Lambda\geq 0$ be a given integer ``length'' scale. All combinations of values of the four indices $i,j,k,l$ belong to exactly one of the following five cases: (1) all indices lie inside an interval of length $\Lambda$, (2) three indices lie within $\Lambda$, the fourth does not, (3) two pairs of indices are separated by more than $\Lambda$, but within each pair, the indices are apart at most by $\Lambda$, (4) one pair lies within $\Lambda$, the other two indices are apart from each other and from the close pair by more than $\Lambda$, (5) all four indices are apart from each other by at least $\Lambda$. Let $n_j$, $j=1,\ldots,5$, be the number of terms in the sum that satisfy cases (1) to (5) above. We have $n_1\leq 4! n\Lambda^3$, $n_2\leq 4! n^2\Lambda^2$, $n_3\leq 4! n^2\Lambda^2$, $n_4\leq 4! n^3\Lambda$, $n_5\leq 4! n^4$. Next, due to the separation of indices and Theorem \ref{corrdeclemma'}, each term of case (5) is of the form
$$
{\mathbb E}[|Y_i| |Y_j| |Y_k| |Y_l|] \leq {\mathbb E}[|Y_i|] {\mathbb E} [|Y_j|] {\mathbb E}[|Y_k|] {\mathbb E}[|Y_l|] +O(\e^{-\gamma'\Lambda})=O(\e^{-\gamma'\Lambda}),
$$ 
as ${\mathbb E}[|Y_i|]=0$ for all $i$. Similarly, each term of cases (4), (3) and (2) have the same upper bound. Each term of case (1) is bounded above by one. We conclude that 
$$
{\mathbb E}[| Z_n-{\mathbb E}[Z_n]|^4] \leq C (n_2+n_3+n_4+n_5)\e^{-\gamma'\Lambda} +n_1.
$$
Choose now $\Lambda =n^\alpha$, with $0<\alpha<2/3$. Then using the above bounds on $n_j$, we have that $n^{-4}{\mathbb E}[| Z_n-{\mathbb E}[Z_n]|^4]$ is summable over $n\geq 1$, i.e., by \fer{u3}, the inequality \fer{u1} holds. This completes the proof of Theorem \ref{theorem2}. \hfill$\blacksquare$

\subsection{Proof of Theorem \ref{prop01'}}

We proceed as in the proof of Theorem \ref{theorem2}. We have 
\begin{eqnarray}
\E[\, \overline{\! X}_{\! n}] &=& \frac1n \sum_{m_1,\ldots,m_n} (m_1+\cdots+ m_n) P(X_1=m_1,\ldots, X_n=m_n)\nonumber\\
&=&\frac1n \sum_{j=1}^n\scalprod{\psi_\s}{T^{j-1} PB^*B\e^{\i\tau K} MP\psi_\s}.
\label{nd1}
\end{eqnarray}
Using that $\lim_{n\rightarrow\infty}\frac1n\sum_{j=0}^{n-1} T^j=\Pi$ and proceeding as in \fer{late1}, we arrive at $\lim_{n\rightarrow\infty}\E[\, \overline{\! X}_{\! n}-\mu_\infty] =0$. By proceeding as in step 2 of the proof of Theorem \ref{theorem2} above, one upgrades this to almost everywhere convergence, $\overline{\! X}_{\! n} \rightarrow\mu_\infty$ a.e.  
The result follows. \hfill $\blacksquare$

\appendix
\section{Ergodic and Riesz projections}
\label{app}

Let $V$ be a power bounded operator on a Hilbert space $\ch$, i.e., such that $\|V^n\|\leq M$ for a constant $M$ independent of $n\in\mathbb N$. Set ${\cal F}=\{\varphi\in \ch\ :\ V\varphi=\varphi\}$ and ${\cal R}={\rm Ran}(1-V)$.

\begin{lem}
\label{lemmaappendix}
We have $\ch={\cal F}+\overline{{\cal R}}$ (closure) with ${\cal F}\cap\overline{{\cal R}}=\{0\}$. Moreover, the projection onto $\cal F$ in this decomposition is the ergodic projection, $\widetilde\Pi=\lim_{N\rightarrow\infty}\frac 1N\sum_{n=1}^N V^n$.
\end{lem}

{\it Proof.\ } Clearly $\frac 1N\sum_{n=1}^N V^n\varphi =\varphi$ for all $N$ and all $\varphi\in\cal F$. Also, $\frac 1N\sum_{n=1}^N V^n(1-V)\psi =\frac 1N V[\psi-V^N\psi]\rightarrow 0$ as $N\rightarrow\infty$, for any $\psi\in \ch$. Thus, if $\varphi\in{\cal F}\cap{\cal R}$ then $\varphi=\frac 1N\sum_{n=1}^N V^n\varphi\rightarrow 0$ as $N\rightarrow\infty$. 
This shows ${\cal F}\cap{\cal R}=\{0\}$. Similarly one shows that ${\cal F}\cap\overline{{\cal R}}=\{0\}$: let $\varphi\in {\cal F}\cap\overline{{\cal R}}$. Then $\varphi=\lim_k\varphi_k$, with $\varphi_k\in {\cal R}$. We have
$$
\varphi=\frac 1N\sum_{n=1}^N V^n\varphi= \frac 1N\sum_{n=1}^N V^n\varphi_k - \frac 1N\sum_{n=1}^N V^n(\varphi_k-\varphi).
$$
The first equality holds since $\varphi\in\cal F$. Since $V$ is power bounded, the norm of the second sum on the r.h.s. is bounded above by $M\|\varphi-\varphi_k\|$ for some $M$ independent of $N$, $k$. The first sum on the r.h.s. converges to zero as $N\rightarrow\infty$, since $\varphi_k\in\cal R$. Thus, upon taking first $k$, then $N$ large enough, we see that $\varphi=0$. This shows that ${\cal F}\cap\overline{{\cal R}} = \{0\}$.

The equality $\ch={\cal F}+\overline{\cal R}$ is equivalent to $\ch=\overline{{\cal F}+{\cal R}}$. We have
\begin{equation*}
{\cal F}^\perp = \overline{ {\rm Ran}(1-V^*)}, \quad 
{\cal R}^\perp = \{\varphi\in\ch\ :\ V^*\varphi=\varphi\}
\end{equation*}
Let $\varphi\in({\cal F}+{\cal R})^\perp$. Then $\varphi\in{\cal F}^\perp \cap {\cal R}^\perp$. However, $V^*$ is power bounded and thus, as above, ${\cal F}^\perp \cap {\cal R}^\perp=\{0\}$. This shows that ${\cal F}+{\cal R}$ is dense in $\ch$. \hfill $\blacksquare$

\medskip

We have $\{\widetilde\Pi=0 \Leftrightarrow {\cal F}=\{0\}\} \Leftrightarrow$ $\{ 1\mbox{\ is not an eigenvalue of $V$}\}$. 
Assume that there is a neighbourhood $U$ of $1$ in the complex plane which does not contain any spectrum of $V$, except possibly the point $1$. Let $\Pi$ be the Riesz spectral projection, $\Pi = \frac{1}{2\pi \i}\oint_\Gamma (z-V)^{-1} {\rm d} z$, where $\Gamma\subset U$ is a simple closed curve encircling $1$. $\Pi$ acts as the identity on $\cal F$, as does $\widetilde\Pi$. Let $\varphi=(1-V)\chi\in\cal R$. Then $\Pi\varphi = (1-V)\Pi\chi$. The operator $(1-V)\Pi$ is the eigen-nilpotent associated to the eigenvalue 1 of $V$. If 1 is a semisimple eigenvalue of $V$, then $\Pi\varphi=0$ for all $\varphi\in\cal R$, and by continuity, $\Pi\overline{\cal R}=\{0\}$, which coincides with the action of $\widetilde\Pi$ on $\overline{\cal R}$. If the eigenvalue 1 is not semisimple (the nilpotent is nonzero), then the action of  $\Pi$ on $\cal R$ does not coincide with that of $\widetilde\Pi$. This shows the following result.
\begin{lem} 
\label{lemma8} 
$\Pi = \widetilde\Pi$ if and only if $\, 1$ is a semisimple eigenvalue of $V$ (or if 1 is not an eigenvalue of $V$, in which case $\Pi = \widetilde\Pi=0$). 
\end{lem}

\begin{lem}
\label{lemma10}
Suppose that $\dim\ch<\infty$. If 1 is an eigenvalue of $V$ then it is semisimple. We thus have $\Pi=\widetilde\Pi$. 
\end{lem}

{\it Proof.\ } Suppose 1 is an eigenvalue of $V$. It suffices to show that $z\mapsto (z-V)^{-1}$ has a simple pole at $z=1$. We have $(z-V)^{-1} = \frac 1{z-1} \widetilde\Pi + (z-\bar V)^{-1}(1-\widetilde\Pi)$, where $\bar V=(1-\widetilde\Pi) V(1-\widetilde\Pi)\upharpoonright_{{\rm Ran}(1-\widetilde\Pi)}$, so we only need to show that $1\notin{\rm spec}(\bar V)$. Suppose that $1\in{\rm spec}(\bar V)$
 and take $\varphi\in {\rm Ran}(1-\widetilde\Pi)$ satisfying $\bar V\varphi=\varphi$, $\|\varphi\|=1$. Since $\widetilde\Pi V=V\widetilde\Pi$ we have $\bar V\varphi = V\varphi$, so $(V^n-1)\varphi=0$ for all $n=1,2,\ldots$ By applying $\frac 1N\sum_{n=1}^N$ and taking $N\rightarrow\infty$, we obtain $(\widetilde\Pi-1)\varphi=0$. This is in contradiction to $\varphi\in {\rm Ran}(1-\widetilde\Pi)$ with $\|\varphi\|=1$.\hfill $\blacksquare$

\section{Logarithmic moment generating function}

It is possible to give general conditions ensuring that the logarithmic moment generating function \fer{001} exists. However, these conditions will be rather abstract (see before \fer{Lambda} below). They amount to knowing that a certain operator ($R_\lambda(\alpha)$ given in \fer{003'} below) has a unique eigenvalue of largest modulus, with corresponding eigenprojection satisfying a non-vanishing overlap condition. We show in this section how perturbation theory can be applied to analyze the spectrum of the operator in question. The formulas established here can be used in the analysis of concrete systems. (An example being the Jaynes-Cummings model of section \ref{example}.)

To calculate the logarithmic moment generating function, \fer{001}, we write 
\begin{eqnarray}
\E[\e^{\alpha(X_1+\cdots+X_n)}] &=& \sum_{m_1,\ldots,m_n}\e^{\alpha(m_1+\cdots +m_n)} P(X_1=m_1,\ldots, X_n=m_n)\nonumber\\
&=& \omega(\e^{\alpha M})^n \scalprod{\psi}{R_\lambda(\alpha)^n\psi},
\label{002'}
\end{eqnarray}
where
\begin{eqnarray}
R_\lambda(\alpha) &=& \sum_m \e^{\alpha m} T_{\{m\}} =P B^*B\e^{\i \tau K} A(\alpha)P \label{003'}\\
A(\alpha) &=& \frac{\e^{\alpha M}}{\omega(\e^{\alpha M})}. 
\label{004}
\end{eqnarray}
The scaling of $A(\alpha)$ by $1/\omega(\e^{\alpha M})$ gives the normalization $R_0(\alpha) = \e^{\i\tau L_\s}$.  The existence of the limit \fer{001} is guaranteed if $R_\lambda(\alpha)$ is diagonalizable and has a unique eigenvalue $\rho_+(\lambda,\alpha)$ of largest modulus, and the corresponding Riesz projection $P_+(\lambda,\alpha)$ satisfies $\scalprod{\psi}{P_+(\lambda,\alpha)\psi}\neq 0$. In this case \fer{001} and \fer{002'} give that
\begin{equation}
\Lambda(\alpha) = \log\omega(\e^{\alpha M}) + \log\rho_+(\lambda,\alpha).
\label{Lambda}
\end{equation}

In concrete examples, it is usually not possible to explicitly evaluate $\rho_+$ (however, it is possible in the example considered in Section \ref{example} !), so perturbation theory is in order. We analyze the analyticity properties of $R_\lambda(\alpha)$ in $\lambda$ and $\alpha$. Let $P_j$ be the orthogonal spectral projections of $M$. For $\alpha\in\mathbb C$ we have 
$$
\|A(\alpha)P\| = \frac{\sqrt{\sum_j \e^{2m_j {\rm Re\, }\alpha} \|P_j\psi_\p\|^2}}{|\sum_j\e^{m_j \alpha }\|P_j\psi_\p\|^2|}.
$$
For $\alpha\in\mathbb R$ there are constants $0<c\leq C<\infty$, independent of $\alpha\in \mathbb R$, s.t. $c\leq \|A(\alpha)P\|\leq C$. This cannot be extended to all complex $\alpha$, since otherwise $A(\alpha)P$ would have to be constant in $\alpha$ by Liouville's theorem of complex analysis. However, if the imaginary part of $\alpha$ is small, then the weighted superposition of the $\|P_j\psi_\p\|^2$ of the denominator is still bounded away from zero. The growth of the numerator and denominator as ${\rm Re}\alpha\rightarrow\pm\infty$ is the same. Thus there is an $\alpha_0>0$ s.t. if $|{\rm Im}\alpha|<\alpha_0$, then $c'\leq \|A(\alpha)P\|\leq C'$ for some $0<c'\leq C'<\infty$.

The operator $R_\lambda(\alpha)$ is holomorphic in $(\lambda, \alpha)\in{\mathbb C}\times\{z\ :\ |{\rm Im}z|<\alpha_0\}$, and has the expansion
\begin{equation}
R_\lambda(\alpha) = \sum_{n\geq 1} \lambda^n R^{(n)}(\alpha),
\label{005}
\end{equation}
with
\begin{equation}
R^{(n)}(\alpha) = \i^n\e^{\i \tau L_\s} \int_0^\tau\d s_1\cdots \int_0^{s_{n-1}}\d s_n PB^*B I(s_n)\cdots I(s_1) A(\alpha)P,
\label{006} 
\end{equation}
and where $I(s) = \e^{\i s L_0} I\e^{-\i s L_0}$. The bound $\|R^{(n)}(\alpha)\|\leq \frac{\tau^n \|I\|^n}{n!} \|B^*B\|\, \|A(\alpha)P\|$ implies that 
\begin{equation}
\sup_{|{\rm Im}\alpha|<\alpha_0} \|R^{(n)}(\alpha)\| \leq C''\frac{\tau^n\|I\|^n}{n!}
\label{007}
\end{equation}
for some $C''<\infty$. Thanks to this bound we can perform perturbation theory in $\lambda$ uniformly in $\alpha$ s.t. $|{\rm Im}\alpha|<\alpha_0$.

\begin{prop}
There are constants $C$, $\lambda_1$, both independent of $\alpha\in\mathbb C$ with $|{\rm Im}\alpha|<\alpha_0$ and of $\tau\geq 0$, s.t. if $|\lambda|<\lambda_1$, then 
\begin{equation}
{\rm dist}\Big( {\rm spec} \big(R_\lambda(\alpha)\big), \ {\rm spec}(\e^{\i\tau L_\s})\Big) \leq  C|\lambda|\tau.
\label{008}
\end{equation} 
Moreover, the group of eigenvalues associated to any two distinct eigenvalues $\e^{\i\tau e}$, $\e^{\i\tau e'}$ of $R_0(\alpha)=\e^{\i\tau L_\s}$ belong to disjoint balls centered at $\e^{\i\tau e}$ and $\e^{\i\tau e'}$.
\end{prop}

A proof is obtained from a straightforward estimate of the resolvent $(R_\lambda(\alpha)-z)^{-1}$ using the Neumann series and the fact that $\|(R_0(\alpha)-z)^{-1}\| = [{\rm dist}(z,{\rm spec}R_0(\alpha))]^{-1}$ since $R_0(\alpha)$ is normal. Due to \fer{007}, the motion of eigenvalues of $R_\lambda(\alpha)$ under variation of $\lambda$  is estimated for $|\lambda|<\lambda_1$, uniformly in $\alpha\in\mathbb C$, $|{\rm Im}\alpha|<\alpha_0$, see \cite{K} Section II \S 3.


{\bf Acknowledgments.} We are grateful to Shannon Starr for pointing out to us the fourth moment method that we use in the proof of step 2 of Theorem \ref{proofthm2}. Our thanks also go to Laurent Bruneau and Alain Joye for many enlightening discussions.  This research has been supported by an NSERC Discovery Grant (Natural Sciences and Engineering Research Council of Canada).

\bibliographystyle{mdpi}
\makeatletter
\renewcommand\@biblabel[1]{#1. }
\makeatother

\end{document}